%% file: main.tex
\PassOptionsToPackage{pdfa}{hyperref}
\documentclass[sigconf, nonacm]{acmart}

\usepackage[a-2b]{pdfx}
\usepackage{xspace}
\usepackage{enumitem}
\usepackage{multirow}
\usepackage{subcaption}
\usepackage{marginnote}
\usepackage[most]{tcolorbox}
\usepackage{xcolor}
\usepackage{soul}
\usepackage{balance}

\usetikzlibrary{matrix,positioning,fit}
\definecolor{websearchColor}{RGB}{217,95,89}   
\definecolor{webcrawlColor}{RGB}{69,123,157}   

\soulregister{\textsc}{1}

\newcommand\vldbdoi{10.14778/3849398.3849417}
\newcommand\vldbpages{5230 - 5243}
\newcommand\vldbvolume{19}
\newcommand\vldbissue{13}
\newcommand\vldbyear{2026}
\newcommand\vldbauthors{\authors}
\newcommand\vldbtitle{\shorttitle} 
\newcommand\vldbavailabilityurl{https://github.com/TheDataStation/pneuma-seeker}
\newcommand\vldbpagestyle{empty} 

\newtcolorbox{revbox}{
    enhanced,
    breakable,
    colback=yellow!25,
    colframe=yellow!25,
    boxrule=0pt,
    arc=0pt,
    left=2pt,
    right=2pt,
    top=2pt,
    bottom=2pt,
    boxsep=0pt
}

\definecolor{shepherdorange}{HTML}{FFD1A9}

\newcommand{\tinyskip}{\vspace{3pt}}
\newcommand{\mypar}[1]{\tinyskip\noindent\textbf{#1.}\xspace}

\newcommand{\UseCasePO}{\hyperref[par:usecase-po]{Use Case: Purchase Order}\xspace}
\newcommand{\ts}{$(\mathcal{T},S)$\xspace}

\newenvironment{myitemize}{%
\begin{itemize}[leftmargin=1em, itemsep=.1em, parsep=.1em, topsep=.1em,
    partopsep=.1em]}
{\end{itemize}}

\newcounter{usecase}

\newcounter{requirement}
\newcommand{\Req}[2]{%
  \refstepcounter{requirement}%
  \noindent\textbf{R\therequirement: #1}\label{#2}
}

\newcommand{\ps}{\textsc{Pneuma-Seeker}\xspace}
\newcommand{\co}{\textsc{Chat-Only}\xspace}
\newcommand{\cs}{\textsc{Chat-Sidebar}\xspace}
\newcommand{\conductor}{\textsc{Conductor}\xspace}
\newcommand{\materializer}{\textsc{Materializer}\xspace}
\newcommand{\retriever}{\textsc{Retriever}\xspace}
\newcommand{\pr}{\textsc{Pneuma-Retriever}\xspace}
\newcommand{\smolagent}{\textsc{smolagents}\xspace}
\newcommand{\dsguru}{\textsc{DS-Guru}\xspace}
\newcommand{\provgraph}{\textsc{ProvenanceGraph}\xspace}
\newcommand{\pneuma}{\textsc{Pneuma}\xspace}
\newcommand{\octopus}{\textsc{Octopus}\xspace}

\newcommand{\duckdb}{\textsc{DuckDB}\xspace}
\newcommand{\dbservice}{\textsc{DBService}\xspace}
\newcommand{\lmservice}{\textsc{LMService}\xspace}

\newcommand{\latentInfoNeed}{$\mathcal{I}^*$\xspace}
\newcommand{\activeInfoNeed}{$\mathcal{I}^+$\xspace}
\newcommand{\outputDocument}{$\mathcal{D}$\xspace}
\newcommand{\techIntervention}{$\Sigma$\xspace}
\newcommand{\tableCollection}{$\mathcal{C}$\xspace}

\begin{document}
\title{Pneuma-Seeker: Relational Reification of Information Needs for Agentic Data Discovery and Preparation}

\input{etc/authors}
\input{camera-ready-sections/0_abstract}

\maketitle

\pagestyle{\vldbpagestyle}
\begingroup\small\noindent\raggedright\textbf{PVLDB Reference Format:}\\
\vldbauthors. \vldbtitle. PVLDB, \vldbvolume(\vldbissue): \vldbpages, \vldbyear.\\
\href{https://doi.org/\vldbdoi}{doi:\vldbdoi}
\endgroup
\begingroup
\renewcommand\thefootnote{}\footnote{\noindent
This work is licensed under the Creative Commons BY-NC-ND 4.0 International License. Visit \url{https://creativecommons.org/licenses/by-nc-nd/4.0/} to view a copy of this license. For any use beyond those covered by this license, obtain permission by emailing \href{mailto:info@vldb.org}{info@vldb.org}. Copyright is held by the owner/author(s). Publication rights licensed to the VLDB Endowment. \\
\raggedright Proceedings of the VLDB Endowment, Vol. \vldbvolume, No. \vldbissue\ %
ISSN 2150-8097. \\
\href{https://doi.org/\vldbdoi}{doi:\vldbdoi} \\
}\addtocounter{footnote}{-1}\endgroup

\ifdefempty{\vldbavailabilityurl}{}{
\vspace{.3cm}
\begingroup\small\noindent\raggedright\textbf{PVLDB Artifact Availability:}\\
The source code, data, and/or other artifacts have been made available at \url{\vldbavailabilityurl}.
\endgroup
}

\input{camera-ready-sections/1_intro}
\input{camera-ready-sections/2_background}
\input{camera-ready-sections/3_reification}
\input{camera-ready-sections/4_context_management}
\input{camera-ready-sections/5_pneuma-seeker}
\input{camera-ready-sections/6_eval}
\input{camera-ready-sections/7_related_work}
\input{camera-ready-sections/8_conclusion}
\input{camera-ready-sections/9_acknowledgement}

\bibliographystyle{ACM-Reference-Format}
\bibliography{main}

\end{document}

%% file: etc/authors.tex
\newcommand{\uchicago}{
  \institution{The University of Chicago}
  \city{Chicago}
  \country{USA}
}

\settopmatter{authorsperrow=4} 

\author{Muhammad Imam Luthfi Balaka}
\email{luthfibalaka@uchicago.edu}
\orcid{0009-0001-5324-7758}
\affiliation{\uchicago}

\author{John Hillesland}
\email{jhillesland@uchicago.edu}
\affiliation{\uchicago}

\author{Kemal Badur}
\email{badur@uchicago.edu}
\affiliation{\uchicago}

\author{Raul Castro Fernandez}
\email{raulcf@uchicago.edu}
\orcid{0000-0001-7675-6080}
\affiliation{\uchicago}

%% file: camera-ready-sections/0_abstract.tex
\begin{abstract}

When faced with data problems, many data workers cannot articulate their information need precisely enough for software to help. Although LLMs interpret natural-language requests, they behave brittly when intent is under-specified, e.g., hallucinating fields, assuming join paths, or producing ungrounded answers.

We present \ps, a system built around a central idea: relational reification. \ps represents a user's evolving information need as a relational schema: a concrete, analysis-ready data model shared between user and system. Rather than answering prompts directly, \ps iteratively refines this schema, then discovers and prepares relevant sources to construct a relation and executable program that compute the answer. \ps employs an LLM-powered agentic architecture with conductor-style planning and macro- and micro-level context management to operate effectively over heterogeneous relational corpora.

We evaluate \ps across multiple domains against state-of-the-art academic and industrial baselines, demonstrating higher answer quality and significant support for iterative refinement. Real-world deployment highlights trust and inspectability as essential requirements for LLM-mediated data systems.

\end{abstract}

%% file: camera-ready-sections/1_intro.tex
\section{Introduction}
\label{sec:intro}
Organizations increasingly make operational, financial, and policy decisions ``in the loop'' with data: analysts, engineers, and domain experts continuously formulate questions, locate relevant datasets, reconcile definitions, and assemble tables or features that can support a decision. In principle, this workflow should be accelerable: a decision maker expresses an \emph{information need}, and a system retrieves and prepares the required data. In practice, however, data work remains hampered by two stubborn stages in the data management lifecycle: i) data discovery, identifying and retrieving documents that could satisfy an information need, and; ii) data preparation, transforming those documents into a usable representation for downstream analysis. These stages resist automation because determining \emph{intent} is expensive: users rarely articulate what they need precisely enough for software systems to help.

Large language models (LLMs) appear to change the landscape. They can interpret natural-language requests, synthesize explanations, and increasingly orchestrate multi-step workflows. This has led to a tempting conclusion: if LLMs can ``understand'' a user request, perhaps they can also discover the right datasets, assemble them, and produce the answer. Our experience building and deploying \ps suggests a more careful view. The bottleneck is not merely that the system cannot execute steps; it is that the user's information need is typically vague, evolving, and difficult to operationalize. Questions are moving targets, and without a concrete target, even very capable models behave brittly, e.g., hallucinating missing fields, assuming join paths, or producing plausible but ungrounded reasoning. What is needed is a mechanism that lets users iteratively converge on a precise intent while giving the system a representation it can reliably act on.

\mypar{Use Case: Purchase Order}
\phantomsection
\label{par:usecase-po}
Consider a procurement office at a large organization trying to cut costs by reducing re-shipping—items that need to be returned and redelivered. A high-level objective (``reduce re-shipping'') is not directly answerable with data. An analyst may hypothesize that certain categories of items cause more downstream friction and ask a question like: \emph{``Do hazardous materials exhibit higher rates of post-purchase order execution friction?''} Yet the organization's data may not contain a clean ``hazardous'' category. Through cross-team conversations the analyst might narrow the concept to radioactive materials; ``execution friction'' may be clarified into measurable signals such as over-shipment, over-invoicing, or forced invoice matching; and the relevant sources may turn out to be spread across multiple systems maintained by different teams. Even after the question becomes crisp, the analyst still faces a discovery and preparation problem: identify which tables contain the relevant signals, determine join keys and grain, reconcile definitions, and produce an analysis-ready relation.

This vignette illustrates three challenges of real data work:

\begin{myitemize}
\item First, information needs are often under-specified initially and become precise only through iteration.

\item Second, the iteration is socially and operationally expensive: it requires coordination and knowledge transfer across teams, tools, and institutional memory. 

\item Third, even after intent stabilizes, retrieval and assembly of the needed data artifacts remains non-trivial.

\end{myitemize}

The ideal system would help users sharpen their intent while reducing cross-team coordination cost, and then would automatically discover and prepare the data needed to answer the now-precise question. Such a system is not a replacement for human communication; instead, it catalyzes convergence toward high-value questions and makes the path from intent to data explicit and verifiable.

This paper investigates whether and how that ideal can be approximated in practice. We present \mbox{\ps}, an LLM-powered system for interactive data discovery and preparation, engineered around \emph{relational reification}: representing a user's evolving information need as a relational schema. Instead of treating the user's question as a single prompt to be answered, \ps treats it as a specification that must be negotiated and made concrete. The system proposes a target relation---attributes, types, and semantics that would constitute an analysis-ready table for the question---and iteratively refines that schema through user feedback. The schema becomes a shared artifact: it is concrete enough for users to critique (\textit{``'hazardous' is too broad; use 'radioactive'''}), and structured enough for the system to operationalize discovery (\textit{``find columns and join paths that can populate these attributes''}).

Relational reification, in the manner we present it here, is new; it changes the nature of the interaction: the user is no longer forced to perfectly describe intent upfront, but can react to a proposed schema, correct it, and progressively expose latent assumptions; the system, in turn, is no longer forced to guess what the answer should look like, but is tasked with producing a relation that matches a specified schema. This aligns the user's language-level intent with the system's data-level operations and enables explainability and verification: once the target schema is explicit, \ps can explain, e.g., which sources contributed to which fields and what transformations were applied. The user can inspect the explanation before trusting the result downstream.

Making relational reification work in practice is itself a systems problem: a schema is only useful if the system can reliably discover sources, populate fields with grounded evidence, and revise the schema as understanding improves. \mbox{\ps} is the system we engineer to make this possible. It decomposes discovery and preparation into specialized subtasks with narrowly scoped contexts—schema refinement, source retrieval, table materialization, join/path reasoning, and program synthesis—orchestrated by a conductor that revises plans based on intermediate results. It also introduces \emph{micro context management}: a deliberate database-probing action that lets agents pull targeted evidence from tabular data (e.g., value distributions, schema samples, join statistics) on demand, rather than relying solely on static sample values as a table's representation. This matters because tabular corpora and enterprise catalogs are large and heterogeneous, and placing all context into a single prompt is often ineffective or infeasible.

We evaluate i) the system's ability to support users in inspecting intermediate states and verifying results during iterative data discovery, ii) the quality of the discovered/prepared relations and the resulting answers, iii) the contribution of context specialization and conductor-style planning relative to alternative designs and baselines, iv) the performance of the system.
We trace observed gains back to \mbox{\ps}'s architectural components, e.g., extending our prior table-discovery \mbox{work~\cite{BalakaPneuma2025}} to reach near-perfect retrieval recall (\mbox{Section~\ref{subsubsec:retrieval_recall}}) and showing that ablating relational reification alone degrades answer quality (\mbox{Section~\ref{subsubsec:value_of_ts}}).
We also report learnings from deploying \ps in a real organizational setting (at The University of Chicago), using observed user feedback on explainability, iteration cost, and trust to distill design requirements for this emerging class of LLM-based data systems and, concretely, the design of \ps.

In summary, this paper makes the following contributions:

\begin{myitemize}
    \item \textbf{Relational Reification.} We introduce the idea of representing an evolving information need as an executable relational schema, allowing users and the system to iteratively negotiate intent.

    \item \textbf{Micro Context Management.} We introduce a deliberate, data-base-probing action for agents to acquire targeted evidence from tabular data (e.g., value distributions, schema samples, join statistics), augmenting static sample values as a table's representation. This action complements the now-standard practice of macro-context management: splitting a task across specialized agents.

    \item \textbf{The \mbox{\ps} System.} We design and implement an agentic system that operationalizes relational reification through context management strategies to discover data, materialize relations, and support provenance-aware iterative refinement.
\end{myitemize}

The rest of the paper is structured as follows. \mbox{Section~\ref{sec:problem_definition}} introduces terminology, the problem statement, and requirements derived from our experience deploying Pneuma at our university. \mbox{Section~\ref{sec:reification_of_info_needs}} presents the idea of \textit{relational reification}, which serves as a nexus between human and machine. \mbox{Section~\ref{sec:sys_design}} then shows how this idea is operationalized through context management techniques and the resulting system architecture. \mbox{Section~\ref{sec:impl_details}} describes the details of the system. We then present evaluation results (\mbox{Section~\ref{sec:eval}}), followed by related work (\mbox{Section~\ref{sec:related_work}}) and conclusions (\mbox{Section~\ref{sec:conclusion}}).

%% file: camera-ready-sections/2_background.tex
\section{Information Needs on Tabular Data}
\label{sec:problem_definition}

This section introduces terminology, presents the problem and its motivation, and states four requirements for a technical solution that we have gathered from deploying the system at our university and gathering feedback from users.

\subsection{Key Terminology}

\mypar{Document} A \emph{document} is a concrete representation of data~\cite{vod}. For example, data about a customer could be represented as a piece of paper, a table in a database, or a PDF; all three are documents. In this paper, we concentrate on tabular documents, such as tables in databases, CSV and Parquet files, and spreadsheets.

\mypar{Information Need} An \emph{information need} characterizes the data required to solve a data-driven task~\cite{vod}. In this work, we assume a one-to-one correspondence between a task and its associated information need, and are interested in finding a document that satisfies that need, i.e., that represents the data required to solve the task that induced it.

In \UseCasePO, the information need characterizes comparable signals of post-purchase order execution friction for specific item categories across procurement, logistics, and finance systems, sufficient to assess which categories contribute disproportionately to re-shipping. This information need is fulfilled by surfacing data that aligns item categories (e.g., radioactive materials) and execution frictions (e.g., over-shipment), allowing the analyst to identify categories that drive re-shipping and ultimately inform interventions toward the high-level objective.

However, even when the task is well-defined, users are often unable to fully and precisely articulate the associated information need at the outset. Relevant variables, subpopulations, or confounding factors may only become apparent after inspecting data, and the adequacy of a data specification may depend on evidence uncovered during analysis. We capture the resulting gap between what is ultimately required and what the user can articulate by distinguishing between \textit{latent} and \textit{active} information needs.

\mypar{Latent Information Need} A \emph{latent information need} \latentInfoNeed denotes the complete and precise characterization of the data required to solve the task. \latentInfoNeed is typically unknown to the user a priori, as it may depend on relevant structure in the data (e.g., subpopulations or distributional properties) that is not evident before exploration.

\mypar{Active Information Need} The \emph{active information need} \activeInfoNeed corresponds to the user's articulation of \latentInfoNeed at a given point in time. \activeInfoNeed reflects their current hypotheses and assumptions and may omit relevant constraints. Through interaction with data and intermediate results, the user may revise \activeInfoNeed to better approximate \latentInfoNeed~\cite{Belkin:1980}.

In \UseCasePO, the analyst may initially specify \activeInfoNeed as comparing friction rates between hazardous and non-hazardous products. After observing that hazardous products span several categories (radioactive, toxin, etc.) with substantially different sample sizes, the analyst may refine \activeInfoNeed to stratify by category and control for sample size, moving closer to \latentInfoNeed.

\subsection{Problem Statement and Motivation}
\label{subec:problem_statement_motivation}

Data discovery is the problem of identifying and retrieving documents that satisfy an information need~\cite{Aurum2018,FernandezDataDI2025}. This problem arises across many domains. For instance, analysts in a retail organization may integrate customer profiles and transaction records to study purchasing behavior~\cite{PredictingCustomerPurchase2024}. Similarly, domain scientists may explore open data repositories (e.g., Chicago Open Data) to analyze behavioral or socioeconomic phenomena~\cite{NumajiriOpenData2024}. More formally:

\emph{Given a collection of tables $\mathcal{C} = \{T_1, \dots, T_n\}$ and a latent information need \latentInfoNeed, the goal is to construct a single document \outputDocument derived from \tableCollection such that \outputDocument fulfills \latentInfoNeed.}

\mypar{Iterative Discovery Process}
Solving the data discovery problem requires gathering documents across heterogeneous \mbox{sources~\cite{DataLakes2023,DataDiscoveryInDataLakes2025}} that are relevant to \mbox{\latentInfoNeed}. A fundamental challenge, however, is that \mbox{\latentInfoNeed} is rarely directly observable by the discovery mechanism. Instead, discovery is mediated through \mbox{\activeInfoNeed}, the user's current (and potentially incomplete) articulation of \mbox{\latentInfoNeed}.

We formalize discovery as an iterative interaction process between a user $\mathcal{O}$ and a discovery mechanism $\Psi$:
\begin{equation*}
    \sigma = \langle (I^+_0, D_0), (I^+_1, D_1), \dots \rangle
\end{equation*}

At each iteration $i$:
\begin{myitemize}
    \item \textbf{Articulation.} $\mathcal{O}$ expresses an active information need $\mathcal{I}_i^+$ that approximates \mbox{\latentInfoNeed} (edge case: $\mathcal{I}_0^+ = \mathcal{I}^*$).
    
    \item \textbf{Synthesis.} $\Psi$ produces a document $D_i = \Psi(\mathcal{I}_i^+, C)$. Importantly, $D_i$ may not fully satisfy $\mathcal{I}_i^+$ due to limitations in either $\mathcal{C}$ or $\Psi$ (e.g., relevant tables may not exist or may not be retrieved).

    \item \textbf{Inspection and Refinement.}
    $\mathcal{O}$ inspects $D_i$ and may subjectively update their understanding of \mbox{\latentInfoNeed}, producing a refined articulation $\mathcal{I}_{i+1}^+$.
\end{myitemize}

\mypar{Towards Convergence}
Because \mbox{\latentInfoNeed} is latent, $\mathcal{I}^*$ and $\mathcal{I}_i^+$ do not inhabit a common, observable metric space. Consequently, $\Psi$ cannot natively compute their distance, nor can it directly optimize for the fulfillment of \mbox{\latentInfoNeed}. Instead, $\Psi$ must act as a mechanism that enables $\mathcal{O}$ to systematically minimize this latent discrepancy.

As a baseline, $\Psi$ must prioritize high-fidelity alignment with $\mathcal{I}_i^+$, i.e., producing $D_i$ that fulfills $\mathcal{I}_i^+$ as much as possible (subject to the available information in $\mathcal{C}$). If this were not the case, $\mathcal{O}$ might not be able to discern whether an inadequate outcome stems from an inherent limitation of information in $\mathcal{C}$ (e.g., relevant tables do not exist) or a failure of $\Psi$ itself (e.g., relevant tables were not retrieved).

Beyond alignment with $\mathcal{I}_i^+$, $\Psi$ acts as a catalyst for convergence, driving $\mathcal{I}_i^+ \to \mathcal{I}^*$ by justifying the synthesis of $D_i$. This interaction leverages the theory of \mbox{\textit{External Cognition}~\cite{1996ScaifeExternalCognition}}, which posits that concrete external representations facilitate \textit{computational offloading} by providing the necessary ``external anchors'' for human reasoning. In this context, $\Psi$ minimizes the cognitive load of the iterative discovery process by providing an interface that exposes the justification for $D_i$ (i.e., the steps $\Psi$ took to produce it). By presenting the current state of discovery ($D_i$ alongside its justification), $\Psi$ establishes the necessary feedback loop for $\mathcal{O}$ to systematically refine $\mathcal{I}_i^+$ into $\mathcal{I}_{i+1}^+$, ultimately converging toward $\mathcal{I}^*$.

\mypar{Scope of the Formalization}
This formalization does not yield convergence guarantees or approximation bounds on $\mathcal{I}_i^+ \to \mathcal{I}^*$: because $\mathcal{I}^*$ is never observable to $\Psi$, no common metric space exists over which such a distance could even be defined, so these guarantees are undefined under this framework, not merely difficult to obtain. Instead, the formalization supplies a vocabulary for justifying concrete design decisions: why $\Psi$ must prioritize high-fidelity alignment with $\mathcal{I}_i^+$ rather than guess at $\mathcal{I}^*$ directly, and why exposing the justification for $D_i$---rather than optimizing some latent objective---is the primary lever available to $\Psi$. Responsibility for closing the gap toward $\mathcal{I}^*$ remains with $\mathcal{O}$; $\Psi$'s role is to make that iteration cheaper and better grounded.

\mypar{Example $\Psi$} One common discovery mechanism is iterative, human-driven workflows that combine manual data discovery, ad hoc integration, and consultation with stakeholders. For example, \UseCasePO requires analysts to manually locate and combine purchase order data and invoicing or execution records, and to consult cross-functional teams to interpret observed patterns. These workflows are communication-heavy and may take weeks.

\mypar{The Opportunity} LLMs demonstrate strong capabilities in interpreting natural-language queries and reasoning over human-specified instructions~\cite{CoT2022,LLMZeroShotReasoners2022,yao2023reactsynergizingreasoningacting}. These capabilities align with the need to interpret \activeInfoNeed \textit{accurately} and reason about actions that would help fulfill it, creating an opportunity to rethink \textit{how} LLMs could reduce the time and effort to address the problem.

\subsection{Solution Requirements}
We outline the requirements of a technical intervention \techIntervention for addressing the problem, informed by user feedback at our university.

\begin{myitemize}
    \item \Req{Alignment Fidelity}{req:correctness}
    Since \mbox{\latentInfoNeed} is unobservable, \mbox{\techIntervention} must ensure high-fidelity alignment with $\mathcal{I}_i^+$: $D_i$ must faithfully follow the constraints articulated in $\mathcal{I}_i^+$ based on the available data in $\mathcal{C}$, allowing the user to isolate data limitations from system failures.
    
    \item \Req{Interpretability}{req:interpretability}
    To ensure that the user can verify (and hence trust) \outputDocument, \techIntervention should justify the derivation of \outputDocument from \tableCollection.

    \item \Req{Refinability}{req:refinability} 
    Since \latentInfoNeed may only be partially specified through \activeInfoNeed, \techIntervention should support mechanisms that allow the user to refine \activeInfoNeed and communicate clarifications or constraints.
    
    \item \Req{Efficiency}{req:efficiency}
    \techIntervention must reduce the time and effort to fulfill an information need compared to current workflows. For example, manual discovery and consultation in procurement scenarios may take weeks and involve multiple stakeholders, whereas \techIntervention should produce \outputDocument within minutes to hours.    
\end{myitemize}

These requirements delineate the space of feasible solutions; approaches that fail any of them are insufficient for the problem as framed. For example:

\begin{myitemize}
    \item Traditional data catalogs~\cite{JahnkeDataCatalogs2023} improve visibility and documentation, but leave users to manually identify tables and refine \activeInfoNeed, failing \textbf{R\ref{req:refinability}} and \textbf{R\ref{req:efficiency}}.

    \item Static dashboards or reporting tools (e.g., \textsc{Tableau}~\cite{tableau}) require users to manually rebuild queries or views to refine \activeInfoNeed, failing \textbf{R\ref{req:refinability}} and often \textbf{R\ref{req:efficiency}}.

\end{myitemize}

Collectively, these examples illustrate that an acceptable technical intervention must not only surface the required data correctly and efficiently, but also support interpretability and refinement of partially-specified information needs.

%% file: camera-ready-sections/3_reification.tex
\section{Relational Reification Mechanism}
\label{sec:reification_of_info_needs}
When an LLM is tasked with directly answering \activeInfoNeed, it must \textit{implicitly} adopt an interpretation of the need and attempt to fulfill it by operating on the underlying tables, entangling the interpretation of \activeInfoNeed with the construction of \outputDocument in a single process. The user has no structured handle on the system's interpretation beyond the final natural-language answer and any associated reasoning traces. Without an explicit representation of this interpretation, the system lacks a grounded notion of progress, and the user cannot easily \textit{understand} \outputDocument well enough to help the system make progress.

For example, in \mbox{\UseCasePO}, the LLM might answer simply \textit{"yes"} or \textit{"no"}, possibly with an explanation of its reasoning. A user seeking to verify correctness may ask follow-up questions, but each response may provide a new explanation, leaving the user uncertain whether to trust it or whether the LLM is hallucinating. Because these explanations are unstructured and inconsistent across queries, the user cannot rely on a uniform method to cross-check answers, making verification ad hoc and error-prone.

We address this by \textbf{reifying} \mbox{\activeInfoNeed} as an explicit relational model. Instead of directly synthesizing \outputDocument, the system first commits to a target model $(\mathcal{T}, S)$, where $\mathcal{T} = \{T_1, \dots, T_k\}$ is a set of derived \emph{views} and $S$ is an executable transformation over $\mathcal{T}$ (e.g., Python code). Each $T_i$ is constructed from \tableCollection via operations, e.g., joins and filters.

Modeling $\mathcal{T}$ as a set of views (instead of a single view) reflects that complex information needs may decompose into multiple semantically distinct relations; for example, in \UseCasePO, different operationalizations of ``hazardous materials'' and different signals of execution friction naturally give rise to multiple views. Collapsing these into one view can obscure modeling assumptions and hinder reuse and inspection. This enforces a separation between \textit{interpretation} and \textit{generation}: the system must make its interpretation of \activeInfoNeed explicit as $(\mathcal{T}, S)$ before producing \outputDocument, decomposing the task into (i) defining $(\mathcal{T}, S)$ and (ii) materializing $\mathcal{T}$, after which executing $S$ yields \outputDocument.

\mypar{Benefits for the User}
Reification externalizes the system's interpretation into concrete, inspectable representations, fulfilling \textbf{R\ref{req:refinability}}. By examining the schemas and sample data of $T$, the user can detect logical misalignment (e.g., missing constraints), allowing them to provide feedback to the system to refine \ts.

\mypar{Benefits for the System}
\ts operationalizes \activeInfoNeed, enabling the system to track progress by checking whether $T$ has been materialized and whether \outputDocument has been produced by executing $S$. It also constrains the system's operation space to actions that support defining \ts, materializing $T$, and executing $S$.

\mypar{Interpretability and Provenance}
While \ts specifies how \outputDocument is constructed, users may want to verify the correctness of $T$ and its derivation. To satisfy \textbf{R\ref{req:interpretability}}, the system records the derivation of $T$ from \tableCollection as a directed acyclic graph (DAG), where nodes represent relational transformations and edges encode data dependencies. This process provides users end-to-end visibility and control over \outputDocument's construction by ensuring the final output is generated by a deterministic script derived directly from the provenance graph.

\mypar{Example: Refining the Relational Model}
Consider \UseCasePO, where the analyst wants to compare post-purchase friction rates for hazardous or radioactive products. The system may initially reify \activeInfoNeed into a single view, $T_{hazardous\_radioactive}$, that combines all hazardous and radioactive products.

Upon inspecting $T_{hazardous\_radioactive}$, the analyst notices that radioactive products are a subset of hazardous products and require separate handling. To enable a more granular analysis, the user refines \activeInfoNeed. The system updates $\mathcal{T}$ to consist of three views: (1) radioactive products, (2) hazardous but non-radioactive products, and (3) non-hazardous products. This refinement ensures that the transformation $S$ produces a final \outputDocument that correctly accounts for subpopulation differences, bringing the \activeInfoNeed closer to \latentInfoNeed.

%% file: camera-ready-sections/4_context_management.tex
\section{Designing a System to Support \ts}
\label{sec:sys_design}
This section presents the design of \mbox{\ps}, an LLM-based system that implements relational reification by reifying \mbox{\activeInfoNeed} as \mbox{\ts}. We discuss two complementary context-management strategies: macro-level, which shapes the overall system architecture, and micro-level, which governs LLM interactions with individual tables.

\begin{figure}[h]
    \centering
    \includegraphics[width=\linewidth]{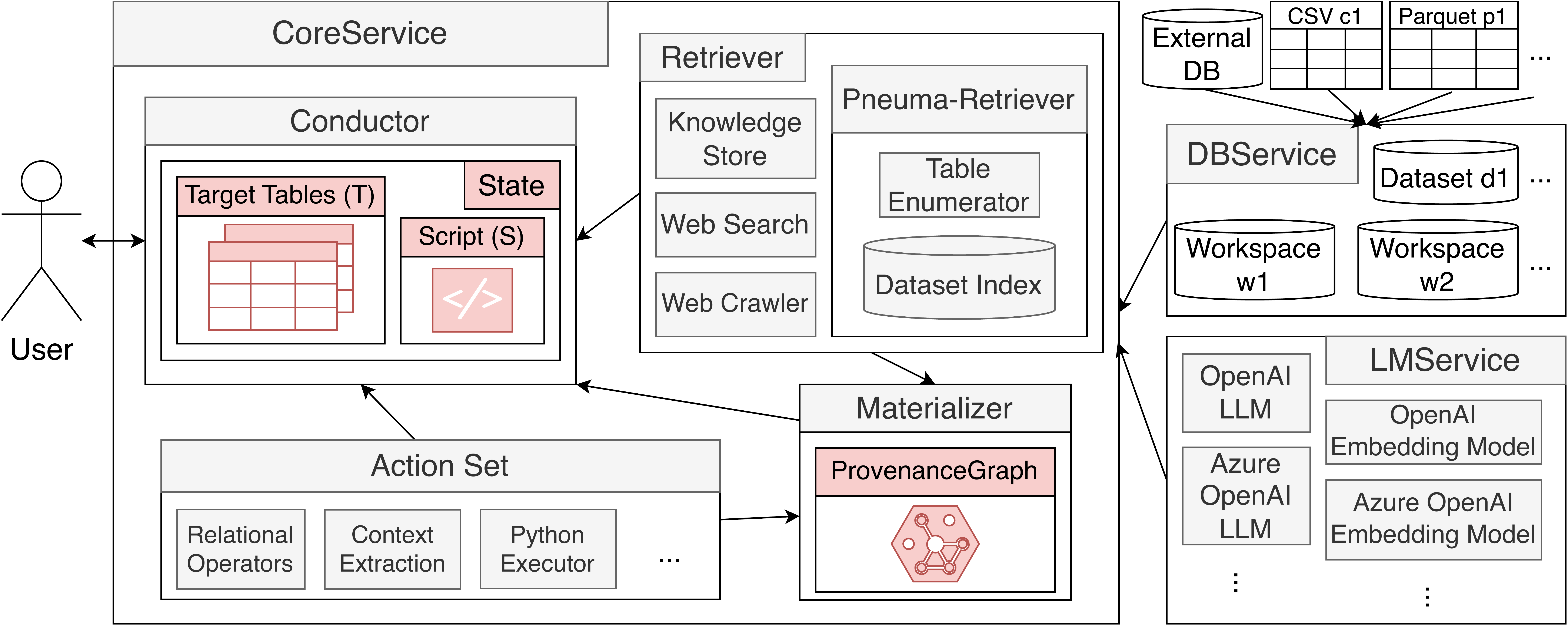}
    \caption{Architecture of \mbox{\ps} ($\rightarrow$: utilized by)}
    \label{fig:architecture}
\end{figure}

\subsection{Macro Context Management}
\label{subsec:macro_context_management}

A central design principle is macro context management: decomposing the workflow into multiple components with scoped responsibilities, thereby reducing the context each component must process. This helps mitigate LLM context window limitations, including the degradation in performance observed with longer contexts~\cite{context-length-hurts-llm,2024LiuLostInTheMiddle}. \ps realizes this principle through a multi-component design (Figure~\ref{fig:architecture}) with three tasks: (1) retrieval, (2) materialization, and (3) orchestration.

Rather than processing all tables in \tableCollection, \ps retrieves a relevant subset by performing tabular data discovery. We encapsulate the retrieval logic in a component called \textbf{\retriever} (Section~\ref{subsec:retriever}). Other parts of the system interact with \retriever only through its interface, which returns relevant tables. This modularization enables extensions (e.g., non-tabular retrieval).

Materializing $\mathcal{T}$ from its specification---rather than the original natural-language phrasing of \activeInfoNeed---constitutes the second main task: given a specification of $\mathcal{T}$ (schemas and column descriptions), materialize $\mathcal{T}$ using tables returned by \retriever. Isolating this functionality enables targeted optimization, e.g., introducing dedicated relational operators to reduce LLM generation errors. This functionality is implemented in \textbf{\materializer} (Section~\ref{subsec:materializer}).

The overall workflow is coordinated by an orchestrator, which implements the relational reification mechanism to transform \activeInfoNeed into \ts. The orchestrator retrieves relevant tables via \retriever to constrain feasible \ts, invokes \materializer to materialize $\mathcal{T}$, and executes $S$ to produce \outputDocument. It is decoupled from the internal implementations of \retriever and \materializer, interacting with them only through their interfaces, and focuses on planning, progress tracking, and user interaction. We define \textbf{\conductor} as the component fulfilling this role (Section~\ref{subsec:conductor}).

These three components define the core service, supported by two additional services: (1) \mbox{\dbservice} (\mbox{Section~\ref{subsec:db_service}}) for persistent storage of \mbox{\ts}, \mbox{\provgraph}, and query execution; and (2) \lmservice (Section~\ref{subsec:lm_service}) for access to language models.

\subsection{Micro Context Management}
\label{subsec:micro_context_management}

Macro-level decomposition alone does not suffice for effective LLM–table interaction. Even after narrowing the search space via \retriever, fully including retrieved tables in context is often infeasible. In \UseCasePO, the purchase order table has 117 columns and 5 million rows; with \verb|o3-2025-04-16|, this would require at least 2.1B tokens---far beyond the 200{,}000-token limit. We often expect larger and multiple tables to be retrieved at once. Conversely, sampling only a few rows can omit critical evidence, leading to incomplete reasoning or errors in constructing \ts.

To address this limitation, \ps introduces a complementary \textit{micro-level context management} strategy. Concretely, micro context management is realized as a structured interaction protocol between the LLM and \dbservice: the LLM is prompted to externalize uncertainty about table contents and may create Python scripts, equipped with APIs to execute SQL queries (part of \mbox{\dbservice}), to issue queries over retrieved tables. Rather than simply supplying table content (i.e., schemas and sample rows), the system actively scaffolds the LLM to \textit{acquire context} on demand by executing targeted probes over tables. This mechanism enables the LLM to retrieve precise values, distributions, and structural properties that are directly relevant to \activeInfoNeed.

For instance, in \mbox{\UseCasePO}, the retrieved PO table lacks a unified ``hazardous'' flag, instead containing columns such as \textit{radioactive} and \textit{toxin}, which may be related. To infer their relationships, the LLM can invoke a script to, e.g., observe the value distributions of these columns. This exploratory probing allows the system to determine exactly which combinations of records may constitute the user's high-level concept before deciding how that subset should be incorporated into $\mathcal{T}$.

Macro context management reduces the contextual load of each component by decomposing the workflow into smaller tasks, while micro context management equips the LLM with an explicit action for fine-grained evidence acquisition within retrieved tables. This explicit support for targeted context extraction directly supports \textbf{R\ref{req:correctness}}; Section~\ref{subsubsec:context_extraction} shows that this action helps produce correct \outputDocument.

As another example, consider the query from the Legal dataset in KramaBench~\cite{lai2025kramabenchbenchmarkaisystems}: \textit{``What is the proportion (round to 3 decimal places) of fraud reporters who lost between \$1--\$500 in 2024?''}. The retriever returns a table whose content interleaves coarse-grained and fine-grained bins; for instance, it includes rows such as ``\$1--\$1{,}000: 624{,}110'' and later ``\$1--\$100: 243{,}174'' and ``\$101--\$200: 114{,}336.'' When only a small set of sample rows is exposed, the LLM assumes that the table provides only coarse-grained aggregates and fails to detect the finer-grained breakdown required to answer the query.

Although the LLM is in principle capable of generating scripts to probe such irregularities, in practice it does not reliably externalize or act on this uncertainty when context extraction is not an explicit, supported action. With micro context management, \ps exposes context extraction as a deliberate action mediated by \dbservice, changing the interaction protocol: the LLM can surface its uncertainty, test it by issuing scripts to probe the table structure (e.g., checking for overlapping bins or finer-grained ranges), surface the relevant rows, and construct \ts accordingly. This enables correct materialization of $\mathcal{T}$ and, in turn, the correct \outputDocument, despite the table's non-ideal organization.

%% file: camera-ready-sections/5_pneuma-seeker.tex
\section{The \ps System}
\label{sec:impl_details}
This section details \textit{how} each main component of \mbox{\ps} performs its task to support the overall workflow: reifying \mbox{\activeInfoNeed} as \mbox{\ts} and executing $S$ to produce \mbox{\outputDocument}.

\mypar{Dynamic Planning}
Rather than following fixed, pre-defined pipelines, \conductor and \materializer construct plans dynamically. Our initial prototypes relied on static pipelines, but we found this approach required costly re-engineering as new use cases necessitated revisiting control-flow decisions (e.g., determining which steps are optional, when to skip or restart actions, and how to integrate new tools). In contrast, dynamic planning allows developers to add new capabilities directly to the components' action spaces, which the planners leverage immediately without altering the overall pipeline, significantly improving extensibility.

Both components operate within bounded planning loops driven by model inference. At each iteration, the planner selects a \emph{sequence} of actions (i.e., agentic tool calls, cf.~ReAct~\cite{yao2023reactsynergizingreasoningacting}), beginning with \textbf{Situational Analysis} to determine subsequent actions in the same iteration. We permit multi-action sequences to balance efficiency and flexibility: a single action per iteration incurs unnecessary model calls and input-token overhead, whereas selecting all actions at once is too rigid because later actions often depend on earlier outcomes (e.g., \conductor defines $\mathcal{T}$ only after gathering evidence from retrieved tables).

\subsection{\conductor}
\label{subsec:conductor}
\conductor orchestrates the overall workflow of the system. Users interact with it through a chat interface. Given a user query, \conductor loops for up to $c$ iterations (default $c=10$). At each iteration, an LLM observes the current state (e.g., \ts and retrieved tables) and outputs a JSON object specifying a sequence of actions to execute. Its action space is as follows:

\begin{myitemize}
    \item \textbf{Situational Analysis.}
    Executed at the beginning of each iteration to analyze the current model \ts and prior actions, and to decide which actions to take in the iteration.

    \item \textbf{\retriever-related Actions.}
    Retrieve the top-$k$ relevant tables from \pr, or search and crawl the web.

    \item \textbf{Context Extraction.}
    Enable micro context management by issuing scripts to probe retrieved tables for relevant evidence (e.g., testing whether specific values exist in certain columns).

    \item \textbf{\ts Manipulation.}
    Update $\mathcal{T}$, $S$, or both, such as adding target views, refining columns, or adjusting transformations.

    \item \textbf{Executor.}
    Execute $S$ over materialized $\mathcal{T}$ to produce \outputDocument.

    \item \textbf{Materializer.}
    Invoke \materializer to materialize $\mathcal{T}$, optionally including textual guidance (e.g., to propagate user feedback).

    \item \textbf{User-Facing Communication.}
    Present results to the user or ask clarifying questions to further refine \ts.
\end{myitemize}

\mypar{Conductor in Action}
Actions are represented as JSON objects containing an action type and corresponding arguments. For example, in the \mbox{\UseCasePO}, the first iteration may produce:

\begin{lstlisting}[belowskip=0pt,aboveskip=0pt]
{"plan":[
  {"action":"Situational Analysis",
   "args":{"message":"No tables yet..."}},
  {"action":"Table Retrieve",
   "args":{"queries":["hazardous..."]}}]}
\end{lstlisting}

In the subsequent iteration, \mbox{\conductor} observes the newly retrieved tables within its context and may combine \textbf{Situational Analysis} with \textbf{Context Extraction} to probe them for specific evidence related to the query. The loop continues until \mbox{\conductor} selects \textbf{User-Facing Communication} or reaches the iteration limit $c$, forcing a final user-facing response.

\subsection{\materializer}
\label{subsec:materializer}
\materializer constructs the views specified in $\mathcal{T}$, taking into account $S$ (if defined by \conductor) and any textual guidance from \conductor. It receives retrieved data as shared context, but may invoke \textbf{Situational Analysis}, \textbf{\retriever-related Actions}, and \textbf{Context Extraction} to reason about intermediate results and acquire additional evidence. Similar to \conductor, \materializer iteratively selects action sequences for up to $m$ iterations (default $m=10$), producing intermediate tables.

\mypar{Table-Construction Actions}
Beyond these shared actions, \materializer has a set of table-construction actions. \materializer favors structured operators over free-form query or code generation, reducing error rates and token usage. Nevertheless, free-form generation is retained as a fallback for transformations that cannot be expressed using structured operators.

\begin{myitemize}
    \item \textbf{Relational Operators.} Common relational operators (e.g., join and projection) are implemented as dedicated actions. The LLM specifies key parameters (e.g., table IDs and join keys), and the application code generates and executes the corresponding queries.

    \item \textbf{Semantic Operators.} We implement several semantic operators. 
    \emph{Semantic join} matches tuples between two tables based on semantic relatedness between textual attributes rather than exact equality, following prior work on embedding-based dataset linking (e.g., \cite{SeepingSemantics2018}) and recent declarative formulations of semantic operators such as LOTUS~\cite{PatelSemanticOperator2024}. We compute embeddings over selected columns and rank candidates using cosine similarity and lightweight syntactic similarity, retaining the top-$p$ matches per left tuple (default $p=1$). 
    \emph{Semantic column generation}, inspired by \textsc{Palimpzest}~\cite{LiuPalimpzest2025}, adds a column to an intermediate table whose values are synthesized by the LLM, conditioned on selected existing columns.

    \item \textbf{Query Executor.}
    Beyond composing relational and semantic operators, \materializer may define custom SQL queries and persist their results as intermediate tables, increasing flexibility without the brittleness of unrestricted code generation.

    \item \textbf{Python Executor.}
    This allows \materializer to generate arbitrary Python code over retrieved and intermediate tables, with access to the DB service APIs (e.g., \verb|execute_query()|). While maximally expressive, it is used only when preceding operators cannot express the required transformation. 
    \end{myitemize}

\provgraph records all transformations performed by \materializer. Nodes correspond to transformations (e.g., joins), while edges encode dependencies between transformations and relations. The graph is constructed incrementally during execution, enabling \mbox{\ps} to expose the derivation of $\mathcal{T}$ via the sidebar (\mbox{Section~\ref{subsec:rq1})}. This is precisely the justification for $D_i$ that our formalization identifies as \mbox{\ps}'s primary lever for supporting the user's own convergence toward \mbox{\latentInfoNeed}. Topologically sorting the graph reconstructs $\mathcal{T}$ as an executable script; appending $S$ yields the end-to-end script that reproduces the generated answer, supporting \mbox{\textbf{R\ref{req:interpretability}}} and \mbox{\textbf{R\ref{req:refinability}}}.

\subsection{\retriever}
\label{subsec:retriever}
\retriever provides a unified interface for all retrieval mechanisms used by \ps. Both \conductor and \materializer may issue up to $q$ retrieval queries in parallel (default $q=3$), each returning top-$k$ items.

The main mechanism is \textbf{\pr}, which retrieves the top-$k$ tables per query (default $k=10$). It builds on \pneuma~\cite{BalakaPneuma2025}, which summarizes table schemas and sample rows for table discovery. Since schema-level summaries miss cell-level signals, we add a content-aware stage inspired by \octopus~\cite{LiOctopus2026}, extracting likely entities from the query, matching them via case-insensitive regex against tables, and scoring matches using damped term frequency $\log(1+\mathrm{tf})$ normalized per keyword. The final score combines content-aware and \pneuma retrieval relevance scores.

\pr also supports enumeration queries over table identifiers via regular expressions (e.g., \texttt{water\_body\_testing\_} \texttt{\textbackslash d\{4\}} after retrieving \verb|water_body_testing_2020|), enabling retrieval when many tables share schemas but differ in attributes such as year or location.

Other mechanisms include: \textbf{Knowledge Store}, which indexes reusable knowledge from user interactions~\cite{park-etal-2021-scalable,liu-etal-2025-user,CaptureKnowledgeUrban2025}; \textbf{Web Search}, which retrieves data via search APIs; and \textbf{Web Crawl}, which fetches user-specified pages when permitted.

\mypar{Extensibility}
\mbox{\retriever} exposes a uniform interface over heterogeneous sources: each result is an \textit{AbstractDocument} with an ID, content, and key--value metadata, where \textit{content} may represent different modalities (e.g., tables or web documents), enabling \mbox{\conductor} and \mbox{\materializer} to operate uniformly across sources. New retrievers can be added by implementing this interface, integrating additional data sources and indexing strategies.

In \mbox{\UseCasePO}, the analyst requests additional information for reporting beyond the retrieved tables. In an actual execution of \mbox{\ps}, \mbox{\conductor} invokes \mbox{\materializer} to augment the target view with non-tabular attributes through Web Search and Web Crawl (\mbox{Figure~\ref{fig:web-augmentation-schemas}}).

\begin{myitemize}
    \item \textbf{Web Search.} After joining the PO and item tables on radioactive-material items, \mbox{\materializer} performs a web search for each supplier company and extracts the \texttt{INDUSTRY\_SECTOR} attribute.

    \item \textbf{Web Crawl.} The user provides a URL template for a radionuclide safety data sheet repository keyed by \texttt{RADIONUCLIDE} (e.g., \texttt{3H}). \mbox{\materializer} crawls one page for each distinct radionuclide and extracts the \texttt{RADIATION\_TYPE} attribute.
\end{myitemize}

In both operations, \mbox{\provgraph} adds a node for each retrieved document upon retrieval. When \mbox{\materializer} integrates the documents through generated Python code, it annotates the code with the documents' IDs; \mbox{\provgraph} detects this annotation and adds the corresponding edge, preserving provenance.

\begin{figure}[h]
\centering
\begin{tikzpicture}[
  every node/.style={font=\ttfamily\tiny, inner sep=1pt},
  cell/.style={draw, minimum height=4mm},
  newWS/.style={fill=websearchColor!25, font=\bfseries\ttfamily\tiny},
  newWC/.style={fill=webcrawlColor!25, font=\bfseries\ttfamily\tiny}
]
\matrix (schema) [matrix of nodes, nodes in empty cells, column sep=-\pgflinewidth,
  nodes={cell}] {
  SUPP\_ID & SUPP\_NAME & |[newWS]| INDUSTRY\_SECTOR & RADIONUCLIDE & |[newWC]| RADIATION\_TYPE & IS\_RADIOACTIVE & IS\_OVERINVOICED \\
};
\node[fit=(schema-1-1)(schema-1-7), inner sep=0pt, outer sep=0pt, below right=0mm and 0mm of schema.north west, anchor=south west, minimum height=4mm, font=\bfseries\sffamily\tiny, fill=gray!25, draw] (hdr) {};
\node[font=\bfseries\sffamily\tiny] at (hdr.center) {PO $\bowtie$ ITEM (augmented)};
\end{tikzpicture}
\caption{Web-Augmented Target View}
\label{fig:web-augmentation-schemas}
\end{figure}

\subsection{\dbservice}
\label{subsec:db_service}
\dbservice provides query execution and persistent storage for system state, including \ts. To support concurrency, each user--chat pair is assigned an isolated workspace database stored as a separate \verb|workspace.db| file using \duckdb~\cite{RaasveldtDuckDB2019}, which localizes failures and transactions. Datasets may be ingested as DuckDB databases (e.g., \verb|procurement.db|) or accessed in external formats such as CSV or Parquet, or external systems such as PostgreSQL via connectors. \dbservice uses DuckDB's direct file querying and \verb|ATTACH| mechanism to unify access across local and external sources.

\subsection{\lmservice}
\label{subsec:lm_service}
\lmservice provides a unified interface to LLMs and embedding models, allowing new providers to be integrated without changes to the core system. We currently support OpenAI and Azure OpenAI endpoints, with Azure used for internal university deployments due to data governance requirements.

%% file: camera-ready-sections/6_eval.tex
\section{Evaluation}
\label{sec:eval}

In this section, we answer the following questions:

\begin{myitemize}
    \item \textbf{RQ1 (User Study)}: Does \mbox{\ps}'s answer justification help users complete tasks?
    \item \textbf{RQ2 (Answer Quality)}: Does \ps fulfill users' information needs?
    \item \textbf{RQ3 (Cost and Scalability)}: What cost--quality and scalability trade-offs does \ps exhibit?
    \item \textbf{RQ4 (Microbenchmarks)}: How do key design choices in \ps shape the trade-offs between answer quality, runtime, memory usage, and retrieval coverage?
\end{myitemize}

\mypar{Datasets} We evaluate \ps on KramaBench~\cite{lai2025kramabenchbenchmarkaisystems}, a benchmark that includes six real-world datasets (Table~\ref{tab:dataset-overview}) spanning scientific, legal, and public domains. We filter the questions to retain only those that require reasoning over tabular data. We additionally preprocess several datasets (e.g., extracting CSVs from XLSX files) to standardize the input format across systems. All questions and processed tables are publicly available in our repository.\footnote{\url{https://github.com/TheDataStation/pneuma-seeker}} 

\setlength{\tabcolsep}{3pt}
\begin{table}[h]
\small
\centering
\caption{Dataset Characteristics and Benchmark Statistics}
\label{tab:dataset-overview}
\begin{tabular}{|c|c|c|c|c|c|}
  \hline
  \textbf{Name} 
  & \textbf{\#Tables} 
  & \textbf{\#Cols}$^\ddagger$ 
  & \textbf{\#Rows}$^\ddagger$ 
  & \textbf{Size (MB)} 
  & \textbf{\#Q} \\
  \hline\hline
  Archeology      & 5       & 16 & 11,289  & 5.90    & 12 \\ \hline
  Astronomy$^*$   & 1,438   & 29 & 1,053   & 313.87 & 4  \\ \hline
  Biomedical      & 24      & 43 & 16,297  & 76.96   & 9  \\ \hline
  Environment$^*$ & 36      & 10 & 9,199   & 29.45   & 17 \\ \hline
  Legal$^*$       & 144     & 3  & 25      & 0.20    & 28 \\ \hline
  Wildfire$^*$    & 20      & 14 & 432,277 & 1040    & 18 \\ \hline
\end{tabular}

\vspace{1ex}
\footnotesize{
$^*$ Filtered to only questions that require tabular data.
$^\ddagger$ Average per table.
}
\end{table}

We choose KramaBench because its questions are non-trivial and predominantly require multi-table integration. In contrast, many existing benchmarks (e.g., MMTU~\cite{xing2025MMTU}, RADAR~\cite{gu2025RADAR}, DataBench~\cite{oses2024databench}) focus on single-table question answering with simpler queries; MultiTabQA~\cite{pal2023multitabqa} supports multi-table integration but with simple questions, and TAG~\cite{biswal2024TAG}, which builds on BIRD~\cite{li2024BIRD}, contains questions with incorrect annotations~\cite{jin2026Text2SQLBroken}. These issues make such benchmarks less suitable for evaluating precise, data-grounded question answering over tables.

\mypar{Baselines} We compare \ps against these baselines:


\begin{myitemize}
    \item \textbf{\dsguru}~\cite{lai2025kramabenchbenchmarkaisystems}: A system that instructs an LLM to decompose a question into subtasks, reason through each step, and synthesize Python code implementing the plan.
    \item \textbf{\smolagent}~\cite{smolagents2024}: An agentic baseline implemented using \smolagent, an open-source agent framework supporting tool calling and code-based reasoning. The agent is provided with tools to execute Python code, list available tables, and inspect schemas and sample rows (up to 5 rows per table).
\end{myitemize}

\mypar{System Setup and Configurations}
Experiments run on a MacBook Air (M4) with 16~GB of RAM and Python 3.12.12, using OpenAI's \verb|o3-2025-04-16| and \verb|text-embedding-3-small| as the LLM and embedding models. To ensure a fair comparison with the baselines, we disable Web Search, Web Crawl, and Knowledge Store.

\subsection{RQ1: User Study}
\label{subsec:rq1}
We recruit 36 participants who interact with data either daily or several times per week. Most participants (36.11\%) are PhD students across various fields (e.g., neuroscience and anthropology), followed by data analysts (11.11\%), engineers (5.56\%), and others.

\begin{figure}[h]
    \centering
    \fbox{\includegraphics[width=\linewidth]{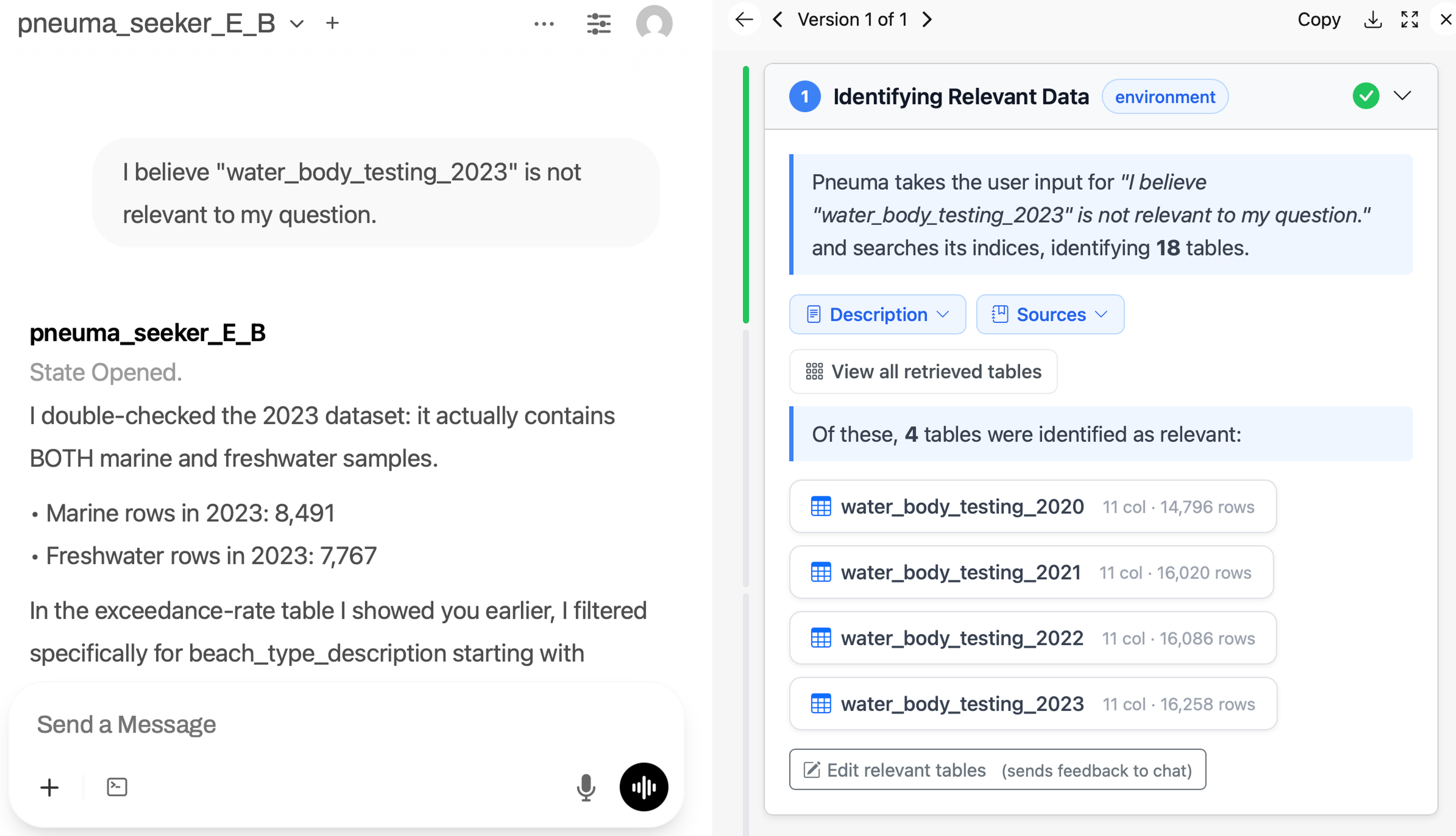}}
    \caption{\mbox{\ps}'s Interface (\mbox{\cs})}
    \label{fig:pneuma-ui}
\end{figure}

We evaluate whether answer justification exposed by \ps supports users in inspecting results during iterative discovery. This targets the proxies for convergence: since \mbox{\latentInfoNeed} is unobservable to the system, we look for observable signals that answer justification helps users close the gap themselves. Concretely, the system provides a sidebar beside the main chat interface (Figure~\ref{fig:pneuma-ui}), which exposes information about the intermediate steps that the system takes to produce an answer. We conduct a within-subjects study, where each participant completes two tasks, one under each condition: \mbox{\cs} (default of \mbox{\ps}) and \mbox{\co}. We counterbalance both condition order and task assignment to mitigate potential order effects.

Each participant is assigned a role and provided with relevant contextual information, then asked to answer two questions from KramaBench. One question is answered using \mbox{\cs}, and the other using \mbox{\co}, with tasks and conditions counterbalanced across participants. The order of conditions is also counterbalanced, such that half of the participants use the \mbox{\cs} first, while the other half begin with \mbox{\co}.
\begin{myitemize}
    \item \textit{``Using the system, determine whether the number of 'frauds and other data' records for 2024 is consistent across datasets for the Metropolitan Area of Miami–Fort Lauderdale–West Palm Beach.''} (role: data analyst at the Federal Trade Commission (FTC))
    \item \textit{``Using the system, determine how many beaches had a higher bacterial exceedance rate in 2013 than in 2012, considering only beaches with samples collected in both years.''} (role: data analyst at the Massachusetts Department of Public Health (DPH))
\end{myitemize}

Participants complete assigned tasks using the system, while also being allowed to inspect and explore artifacts beyond directly prompting it. For instance, they may download relevant tables from the sidebar and independently cross-check results. After exploration, participants rate: (1) their confidence in the correctness of the system's answer and (2) their readiness to act on the result (e.g., to refine information needs or make decisions).
These measures serve as observable proxies for convergence toward \mbox{\latentInfoNeed} under our formalization, since the distance to \mbox{\latentInfoNeed} itself cannot be measured directly.
Responses are collected using a 5-point Likert \mbox{scale~\cite{likert1932technique}}. Participants are also asked to think aloud during interactions with both conditions and describe how they inspect answers. Based on these observations, we classify inspection effort as: (1) low (relying only on chat responses without, e.g., checking calculations); (2) medium (inspecting answer justification through the sidebar); or (3) high (performing manual audits such as downloading the tables and performing calculations).

\mbox{Table~\ref{tab:user-ratings}} shows \mbox{\cs} achieves higher median confidence and readiness-to-act scores than \mbox{\co}. We further analyze the intra-condition rank correlation (Kendall's $\tau_b$) between inspection effort and these ratings. In \mbox{\co}, both correlations are significant and positive (e.g., $\tau_b = .338, p = .0292$ for confidence): users need increasingly manual inspection --- recalculating statistics, cross-checking source tables --- before feeling confident or ready to act. In \mbox{\cs}, both correlations become non-significant (e.g., $\tau_b = .063, p = .6621$ for confidence): confidence and readiness to act no longer depend on how much users manually verify.
This indicates that exposing answer justification through the sidebar substitutes for manual verification as a basis for confidence and readiness to act --- consistent with the role of justification in our formalization, where $\Psi$ cannot optimize for \mbox{\latentInfoNeed} directly and instead supports the user's own convergence by justifying $D_i$, reducing the verification effort $\mathcal{O}$ must supply.

\setlength{\tabcolsep}{2pt}
\begin{table}[h]
\centering
\caption{Ratings (5-Point Likert) and $\tau_b$ vs. Inspection Effort}
\label{tab:user-ratings}
\begin{tabular}{|l|c|c|c|c|c|c|}
  \hline
  \multicolumn{1}{|c|}{\multirow[c]{2}{*}{\textbf{Measure}}}
& \multicolumn{3}{c|}{\textbf{\co}} 
& \multicolumn{3}{c|}{\textbf{\cs}} \\
\cline{2-7}
& \textbf{Median} & \textbf{$\tau_b$} & \textbf{$p$}
& \textbf{Median} & \textbf{$\tau_b$} & \textbf{$p$} \\
\hline\hline
  Confidence       & 3.50 & 0.338 & \textbf{0.0292} & 4.00 & 0.063 & 0.6621 \\ \hline
  Readiness to Act & 2.50 & 0.398 & \textbf{0.0095} & 3.00 & 0.082 & 0.5687 \\ \hline
\end{tabular}
\end{table}

To further contextualize the size of this improvement, we highlight qualitative feedback from participants:

\begin{myitemize}
    \item \textit{``So I could kind of like reason through the steps... I can see the source data... I can now see how this leads to the final output... So I could kind of like reason through... so now I'm super confident.''}

    \item \textit{``...this interface makes it really easy for me to follow what’s going on... for Miami-Fort Lauderdale, it’s 171 for number reports, and then it’s 21,793 for reports per 100K, and that’s wrong...''}

    \item \textit{``The code explanation is very helpful in terms of you can kind of matches what is the variable in the code and what it correspond to the columns in table... [The system] shows me where the data comes from, where the how it is being processed...''}
\end{myitemize}

\subsection{RQ2: Answer Quality}
\label{subsec:rq2}
While RQ1 evaluates the justification lever, RQ2 evaluates the other lever identified in our formalization: high-fidelity alignment with the current active information need $\mathcal{I}_i^+$, the baseline \ps must satisfy for that justification to be meaningful. We measure whether the system answers the questions correctly, i.e., to what extent \ps meets \textbf{R1}. Concretely, we treat each question in KramaBench as $\mathcal{I}_i^+$, and define \textbf{answer quality} as the proportion of correctly answered questions across the benchmark. For questions whose answers are sets or lists of values, we use the F1 score to account for partial correctness.

\begin{figure*}[t]
    \centering
    \includegraphics[width=\linewidth]{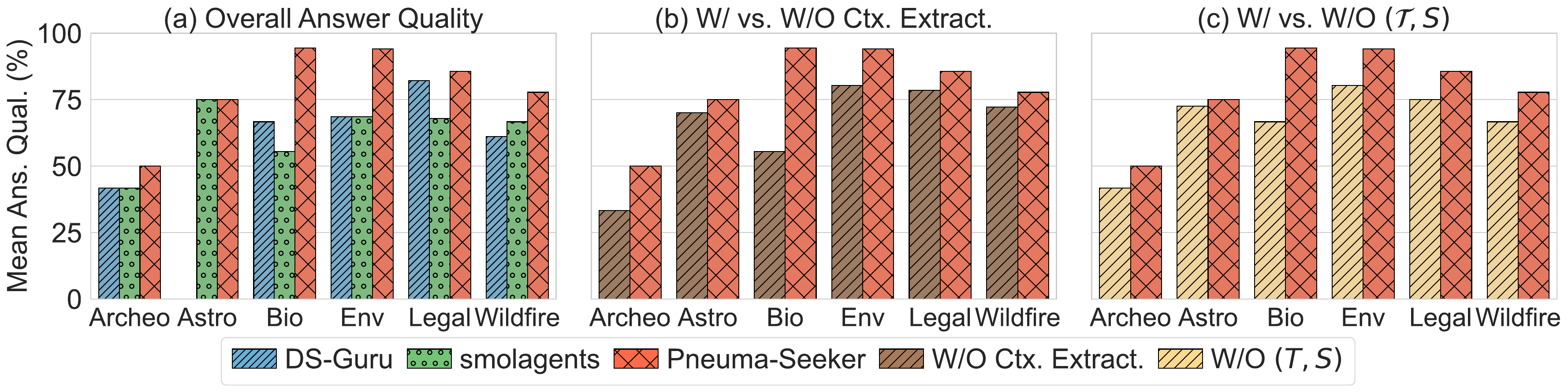}
    \caption{Comparison of Answer Quality}
    \label{fig:answer_quality_aio}
\end{figure*}

\subsubsection{Overall Answer Quality}
\ps achieves higher answer quality than the baselines (Figure~\ref{fig:answer_quality_aio}(a)), on par with \smolagent on the Astronomy dataset. The gap is most pronounced on Biomedical and Environment: on Biomedical, \ps attains an answer quality of $94.44\%$, exceeding \dsguru and \smolagent by $27.77$ and $38.88$ percentage points, respectively.

For \dsguru, we limit sampled rows per table to $5$ on Biomedical to fit the 200{,}000-token context window. On Astronomy, even limiting to $1$ row per table still requires 986{,}955 tokens, far exceeding the limit, so we omit \dsguru on Astronomy.

\subsubsection{Value of Context Extraction}
\label{subsubsec:context_extraction}
We demonstrate the importance of micro context management for answer quality by ablating Context Extraction---the action implementing micro context management in \ps---and comparing the answer quality of \ps with and without this action.

As shown in Figure~\ref{fig:answer_quality_aio}(b), enabling Context Extraction improves answer quality on $4/5$ datasets. Beyond the example in Section~\ref{subsec:micro_context_management}, we highlight a representative failure example that arises without Context Extraction. Consider this query from the Archeology dataset: \textit{``What is the average latitude of capital cities? If there are more than one capital in a country, only count the latitude of the capital with the largest population. Round your answer to 4 decimal places.''} The relevant table is \textit{worldcities} with the following schema: \texttt{[city,lat,lng,capital,...]}

Without Context Extraction, \conductor instantiates a filter predicate from the surface form of the column name \texttt{capital}, assuming any non-empty value denotes a national capital. This leads to an incorrect predicate of the form \texttt{capital IS NOT NULL AND capital <> ''}, which conflates national capitals with administrative and minor capitals. The query executes successfully, but it computes the statistic over a semantically incorrect population.

With Context Extraction enabled, \conductor inspects the empirical distribution of the \texttt{capital} column before defining \ts. Discovering that \texttt{capital} contains {\texttt{primary}, \texttt{admin}, \texttt{minor}, \texttt{NULL}}, where \textit{primary} uniquely identifies national capitals, grounds the subsequent query construction. This yields the correct predicate \texttt{capital = 'primary'} prior to the per-country maximum population selection and aggregation.

Overall, Context Extraction explicitly nudges the LLM to make data-grounded decisions by inspecting actual table values. This directly conditions downstream processes (e.g., defining \ts) on empirical observations rather than implicit assumptions.

\subsubsection{Value of \mbox{\ts}}
\label{subsubsec:value_of_ts}
Beyond helping transform \activeInfoNeed into \latentInfoNeed, \ts also impacts answer quality because it grounds the system's behavior. We compare the answer quality of \ps with and without \ts: without it, \conductor no longer defines or materializes \ts, and instead relies on Context Extraction to directly synthesize an answer.

\ps with \ts consistently achieves higher answer quality than without it (Figure~\ref{fig:answer_quality_aio}(c)). Consider a representative failure on the Legal dataset, which requires data across all U.S. states: \textit{``How many total Identity Theft reports were there in 2024 from cross-state Metropolitan Statistical Areas?''}

Without \ts, the system attempts to generate an answer by reasoning over only the state-level tables it has retrieved, resulting in incomplete coverage. Some states are not included, leading to an underestimated total of 114,392 reports in our experiment.

With \ts, \conductor defines $\mathcal{T}$ as a single table \verb|identity_| \verb|theft_msa_reports| with schema: \verb|[state, metropolitan_area,| \verb|num_of_reports]|, and $S$ specifies the operations needed to consolidate, deduplicate, and sum the reports. \conductor invokes \materializer with the note to populate \verb|identity_theft_| \verb|msa_reports| by UNION-ing all tables whose names match \verb|state_msa_identity_| \verb|theft_data_*|. \materializer ensures that all relevant tables are retrieved, so no state is omitted. Execution of $S$ produces the correct total of 243,377 reports as intended.

Even without \ts, the system can still enumerate tables, but \ts explicitly encodes cues that guide correct materialization, ensuring that all relevant tables contribute to the final answer.

\subsection{RQ3: Cost and Scalability}
\label{subsec:rq3}
We compare the cost and scalability of \ps against the baselines to study how \ps meets \textbf{R4}. Specifically, we measure: (1) monetary cost for LLM inference, (2) runtime scalability as dataset size increases, and (3) the overall runtime of the system across datasets.

\subsubsection{Cost--Quality Trade-offs}
\begin{figure}[h]
    \centering
    \includegraphics[width=\linewidth]{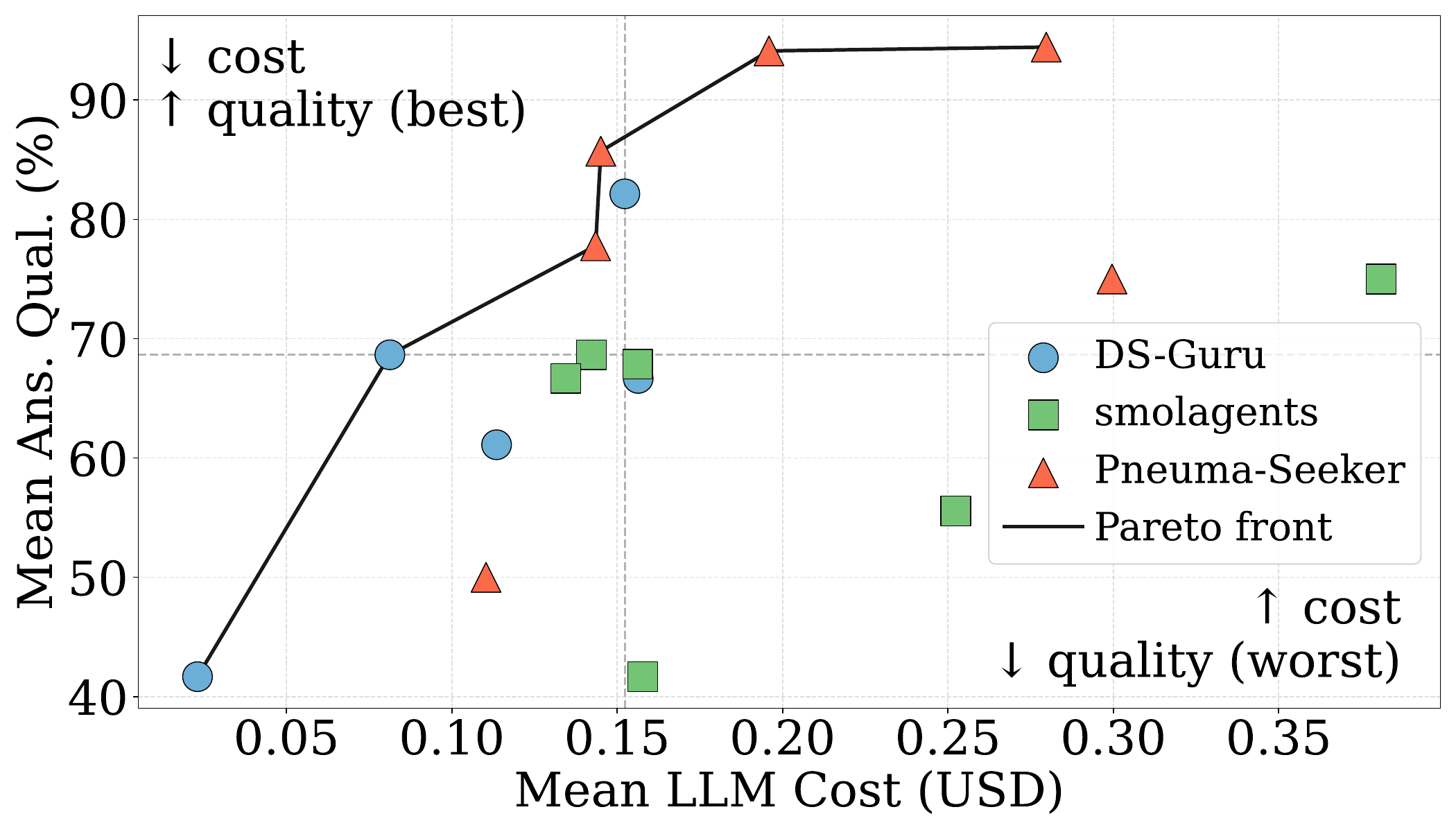}
    \caption{LLM Cost-Answer Quality Trade-Offs}
    \label{fig:rq2_pareto_quality_vs_cost}
\end{figure}

We derive cost from input and output token usage at \texttt{o3-2025-04-16} pricing\footnote{As of May 24, 2026, \texttt{o3-2025-04-16} costs \$2 and \$8 per 1M input and output tokens.} and plot average LLM cost against answer quality in Figure~\ref{fig:rq2_pareto_quality_vs_cost}. \ps occupies most of the Pareto-optimal frontier, achieving state-of-the-art cost--quality efficiency.

\begin{figure*}[t]
    \centering
    \hfill
    \begin{subfigure}[t]{0.32\linewidth}
        \centering
        \includegraphics[width=\linewidth]{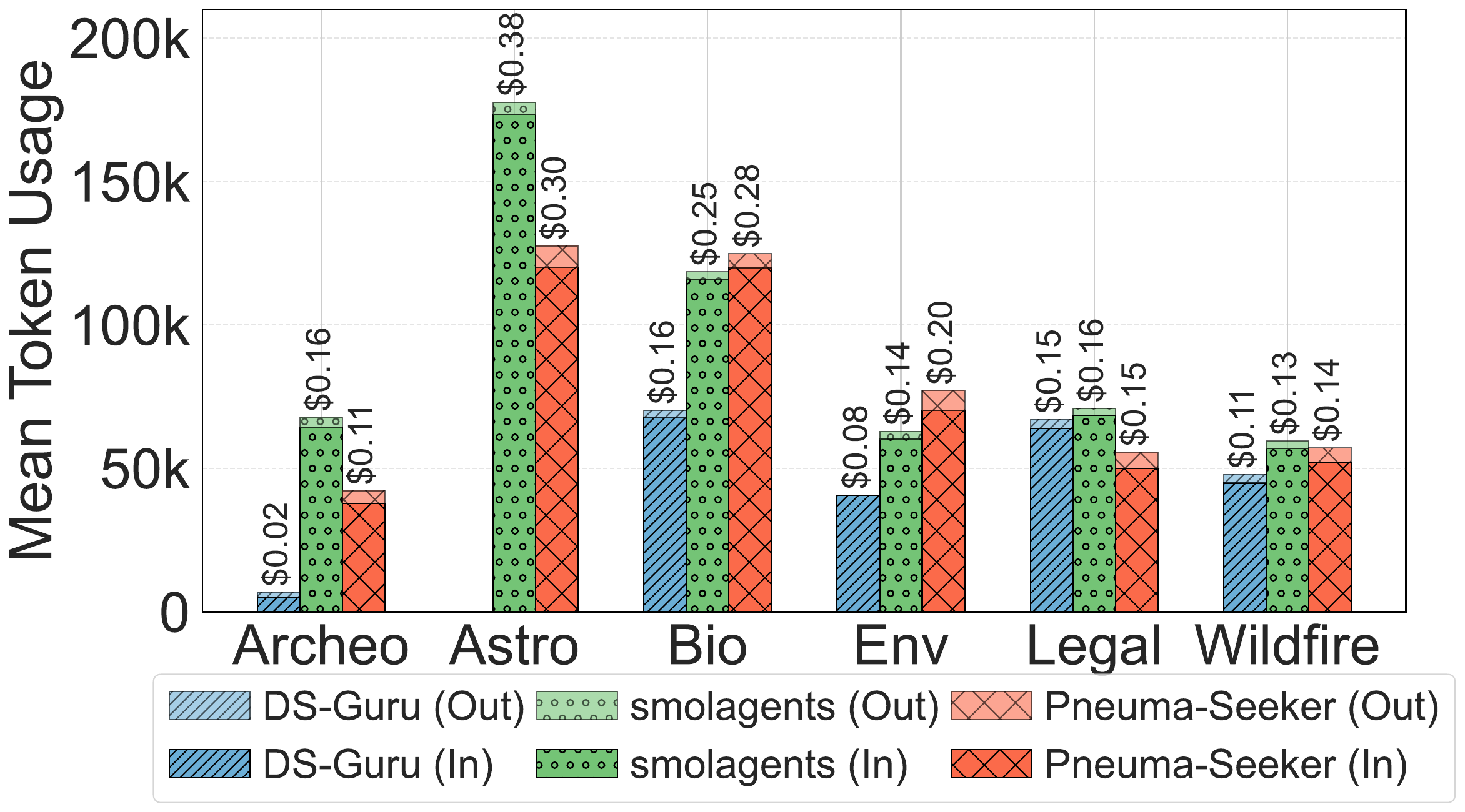}
        \caption{Mean Input and Output Token Usage}
        \label{fig:rq2_token_usage}
    \end{subfigure}
    \hfill
    \begin{subfigure}[t]{0.32\linewidth}
        \centering
        \includegraphics[width=\linewidth]{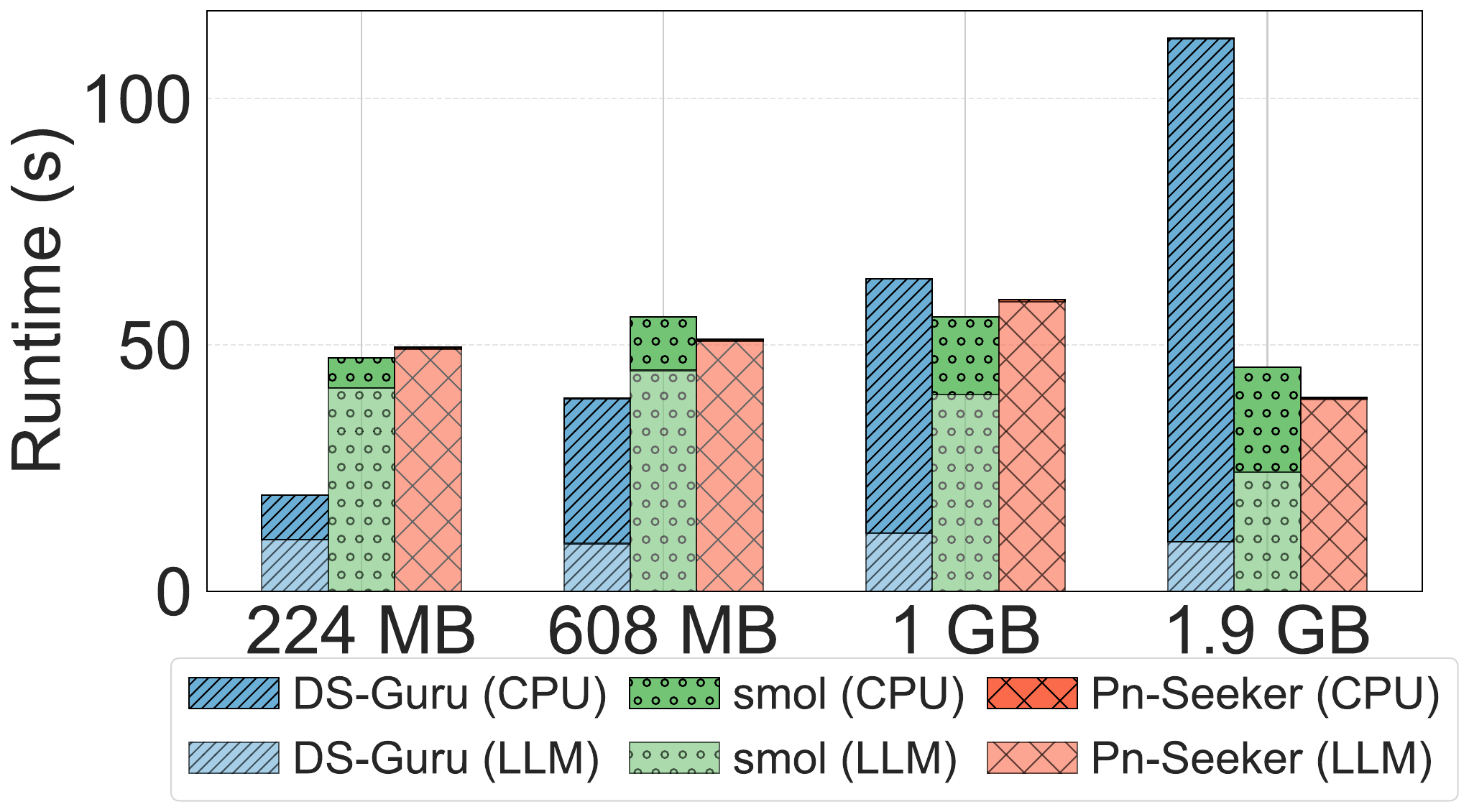}
        \caption{Scalability of LLM \& Non-LLM Runtime}
        \label{fig:rq2_scalability_runtime}
    \end{subfigure}
    \hfill
    \begin{subfigure}[t]{0.33\linewidth}
        \centering
        \includegraphics[width=\linewidth]{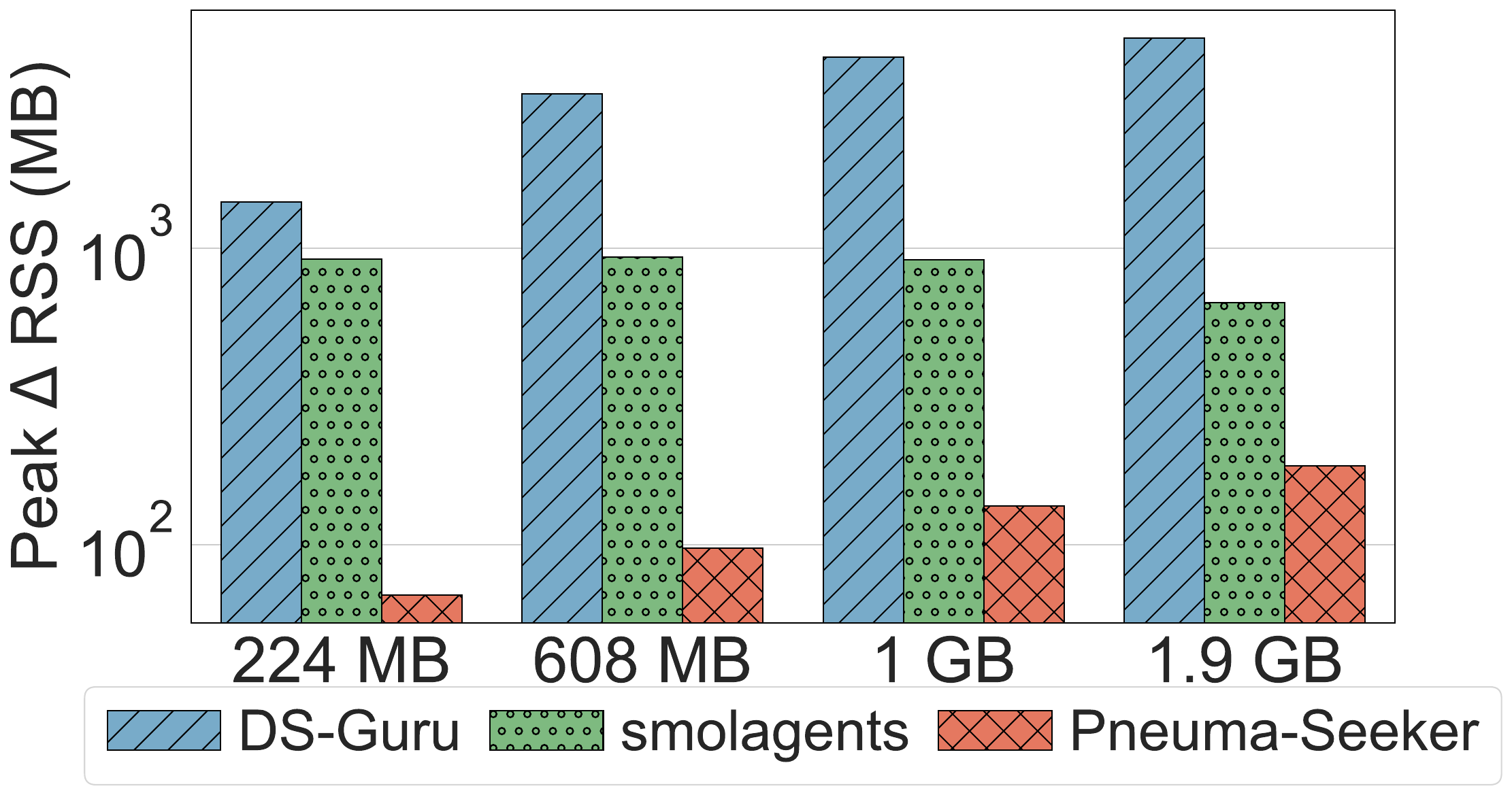}
        \caption{Scalability of Memory Usage}
        \label{fig:rq2_scalability_memory}
    \end{subfigure}

    \caption{Cost and Scalability Analysis}
    \label{fig:rq2_combined}
\end{figure*}

We also report the average input and output token usage for LLM inference across datasets (Figure~\ref{fig:rq2_token_usage}): \dsguru consistently uses the fewest tokens, while \ps's token usage is similar to \smolagent's and lower on 4 out of 6 datasets.

A major reason for this cost–quality trade-off is context management at both retrieval and reasoning levels. On the cost side, narrowing the candidate space via \mbox{\retriever} reduces token consumption, and \mbox{\materializer}’s preference for structured relational operators further limits verbose free-form code generation. On the quality side, micro-level context extraction enables \mbox{\ps} to incorporate salient evidence beyond schemas and sampled table rows, improving answer quality.

\subsubsection{Scalability}
\label{subsubsec:scalability}
We examine \ps's performance profile as dataset size increases, evaluating runtime scalability using a subset of the procurement dataset introduced in Section~\ref{subsec:rq2}. Specifically, we use two tables corresponding to purchase order lines and item details, and vary the number of rows to construct datasets of increasing size: 224~MB, 608~MB, 1~GB, and 1.9~GB. The query remains fixed: \textit{``What is the grand total amount for purchase order lines that include green products?''}

\ps is comparable to \smolagent, while \dsguru eventually overtakes both systems at the 1~GB scale (Figure~\ref{fig:rq2_scalability_runtime}). At that scale, \ps spends 58.74 and 0.37 seconds on LLM inference and non-LLM processing, respectively, compared to 39.94 and 15.75 for \smolagent and 11.73 and 51.71 for \dsguru.

The runtime bottleneck of \smolagent gradually shifts toward non-LLM processing as dataset size increases, similar to but less extreme than \dsguru. For \ps, in contrast, runtime remains dominated by LLM inference and does not grow visibly relative to non-LLM processing time. This highlights the importance of database-backed execution: even when memory pressure is partially controlled via batching, non-LLM processing overhead can dominate runtime when data are loaded and processed in memory.

\subsubsection{Runtime}
\begin{figure*}[t]
    \centering
    \begin{subfigure}[t]{0.30\linewidth}
        \centering
        \includegraphics[width=\linewidth]{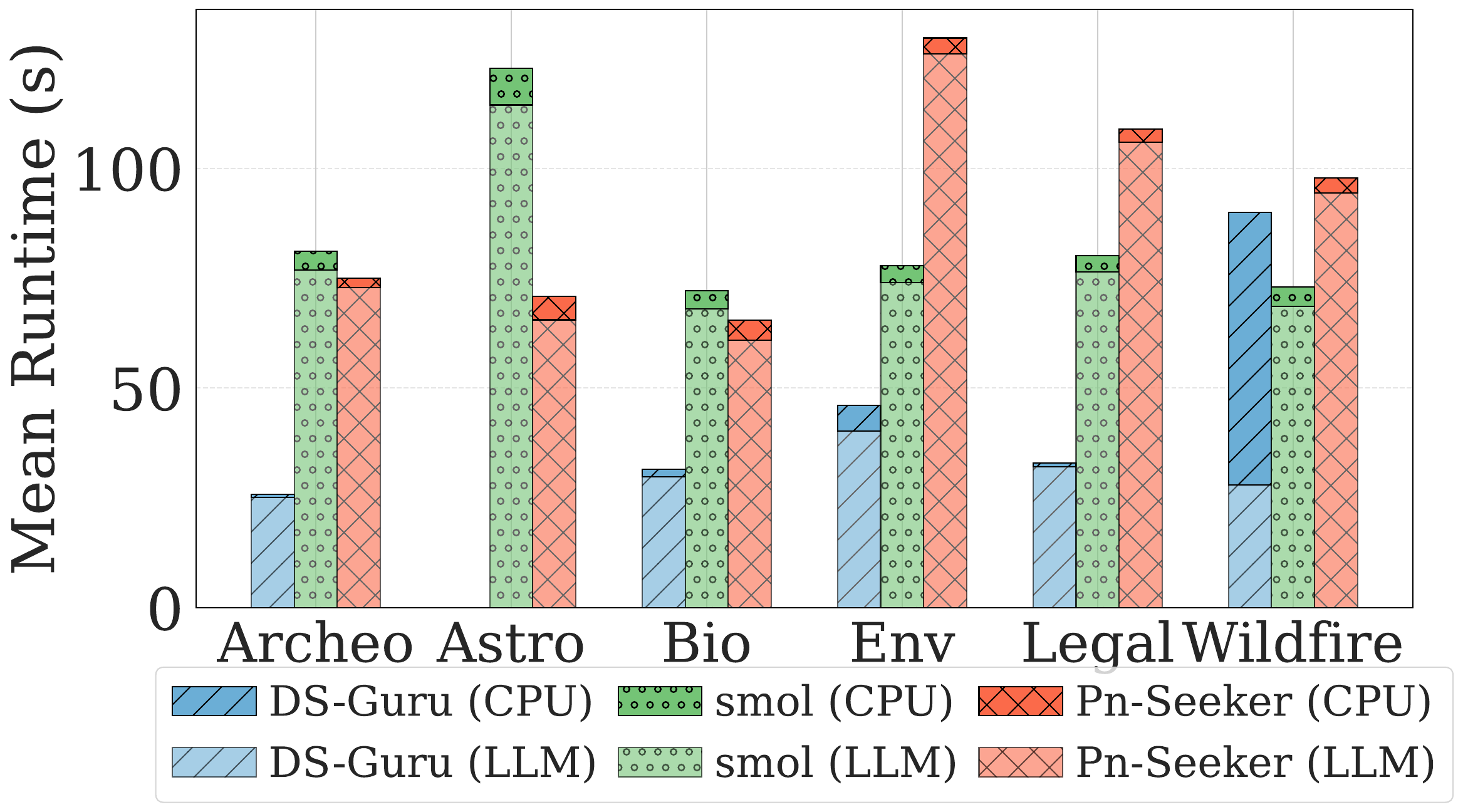}
        \caption{LLM and non-LLM Runtime Breakdown}
        \label{tab:runtime-breakdown}
    \end{subfigure}
    \hfill
    \begin{subfigure}[t]{0.30\linewidth}
        \centering
        \includegraphics[width=\linewidth]{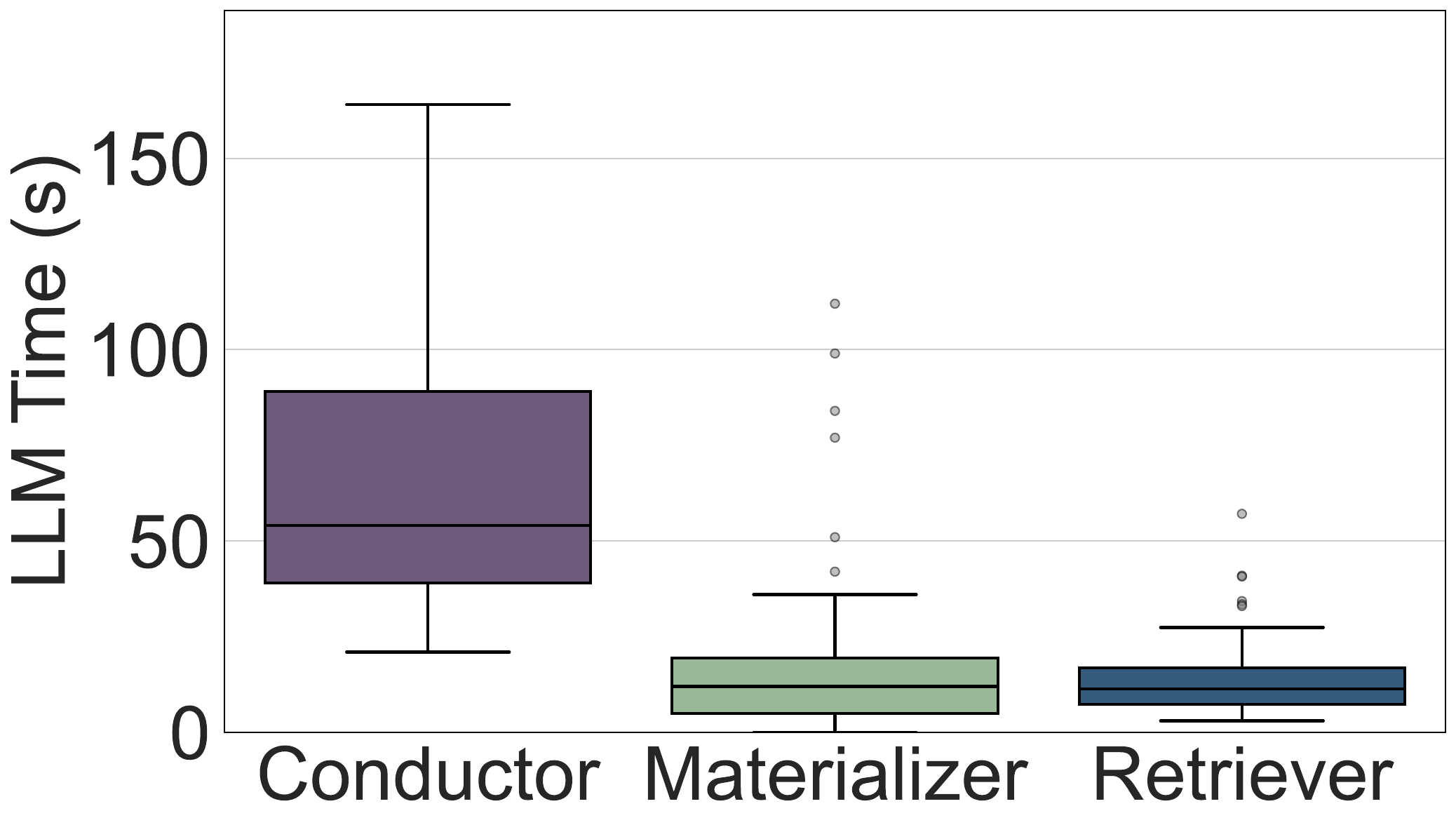}
        \caption{\mbox{\ps}'s LLM Inference}
        \label{fig:llm_time_breakdown}
    \end{subfigure}
    \hfill
    \begin{subfigure}[t]{0.30\linewidth}
        \centering
        \includegraphics[width=\linewidth]{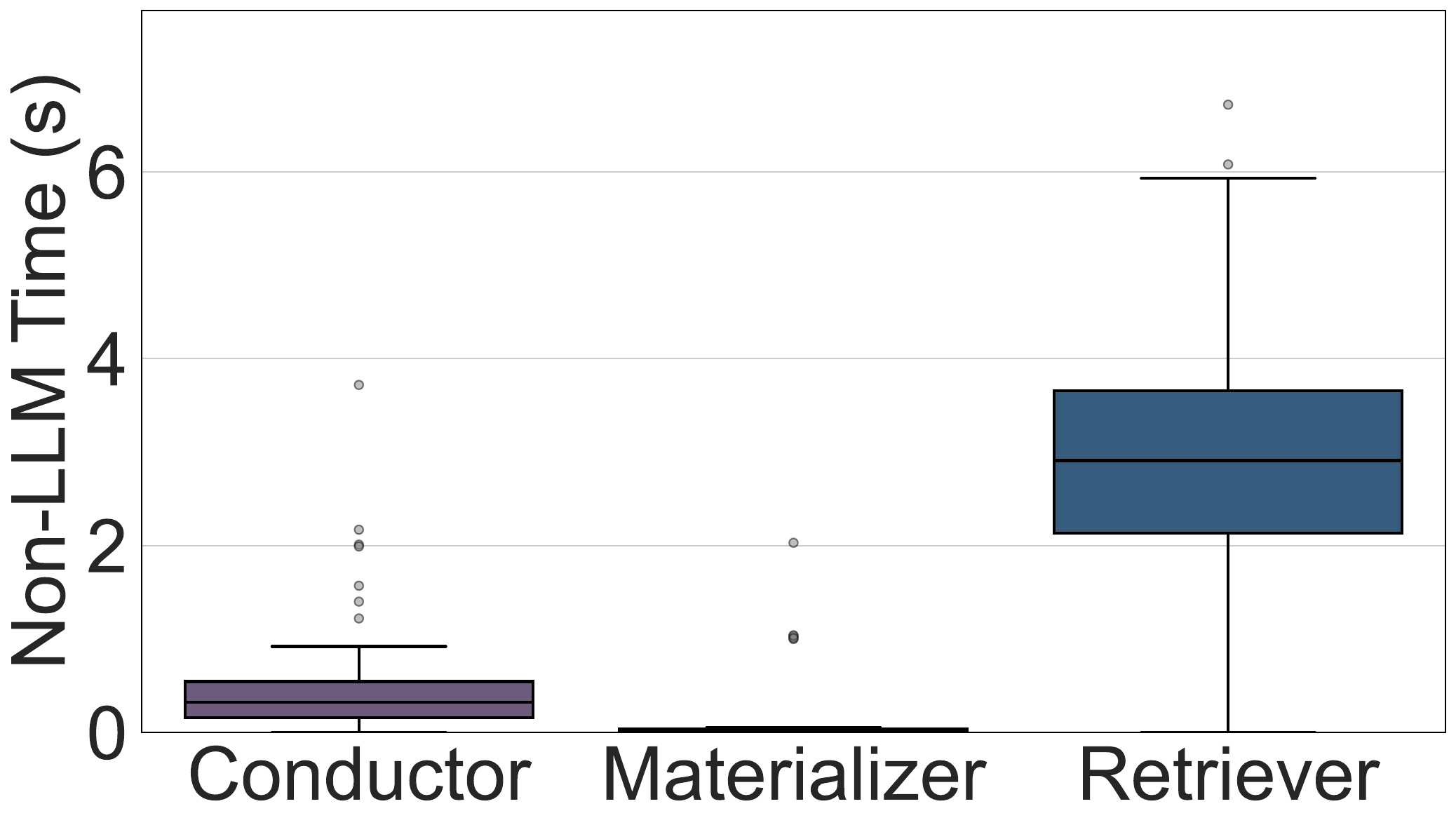}
        \caption{\mbox{\ps}'s Non-LLM Execution}
        \label{fig:non_llm_time_breakdown}
    \end{subfigure}
    \caption{Runtime Breakdown of Systems and \mbox{\ps} in Particular}
    \label{fig:ps_runtime_breakdown}
\end{figure*}

While \ps incurs higher runtime than \dsguru on several datasets, this overhead primarily reflects the additional structured processing introduced by \ts, not inefficiencies in data access or execution. \ps takes longer than \dsguru but is comparable to \smolagent (Figure~\ref{tab:runtime-breakdown}): faster on three datasets and slower on three. For example, on the Biomedical dataset, \ps spends 60.94 seconds on LLM inference on average ($\sigma=35.51$), versus 68.05 ($\sigma=49.07$) for \smolagent and 29.86 ($\sigma=11.99$) for \dsguru.

Across all systems, non-LLM running time is substantially lower than LLM inference time: \ps, \smolagent, and \dsguru spend 4.55, 4.11, and 1.62 seconds on average, respectively (standard deviations: 1.56, 1.02, and 1.68).

For both \ps and \smolagent, LLM inference dominates runtime, whereas for \dsguru, the bottleneck shifts to non-LLM processing on the Wildfire dataset, where \dsguru spends 62.10 seconds ($\sigma=9.79$) on non-LLM time on average, compared to 27.92 seconds ($\sigma=13.52$) on LLM inference, because it loads all rows of all tables into memory. \smolagent, in contrast, implicitly prompts the LLM to process data in batches, while \ps relies on database-backed execution and loads only sample rows into memory. This design substantially reduces non-LLM processing overhead and contributes to the scalability behavior observed in Figure~\ref{fig:rq2_scalability_runtime}; as we further analyze in Section~\ref{subsubsec:memory_usage_runtime_memory_tradeoffs}, this overhead manifests primarily as increased runtime rather than increased memory footprint, reflecting a deliberate design trade-off in \ps.

\mypar{Runtime Breakdown of \ps}
Focusing on the runtime of \ps, we break down both the LLM and non-LLM time for the three main components: \conductor, \materializer, and \retriever. The bottleneck for LLM time (\mbox{Figure~\ref{fig:llm_time_breakdown}}) is \mbox{\conductor}, while for non-LLM time (\mbox{Figure~\ref{fig:non_llm_time_breakdown}}) it is \mbox{\retriever}.

\subsection{RQ4: Microbenchmark}
\label{subsec:rq4}
We complement RQ2 and RQ3 with the analysis of (i) the answer quality on different LLMs, (ii) memory behavior and runtime--memory trade-offs, and (iii) the retrieval recall of \ps.

\subsubsection{Answer Quality Across Different LLMs}
\begin{figure}[h]
    \centering
    \includegraphics[width=\linewidth]{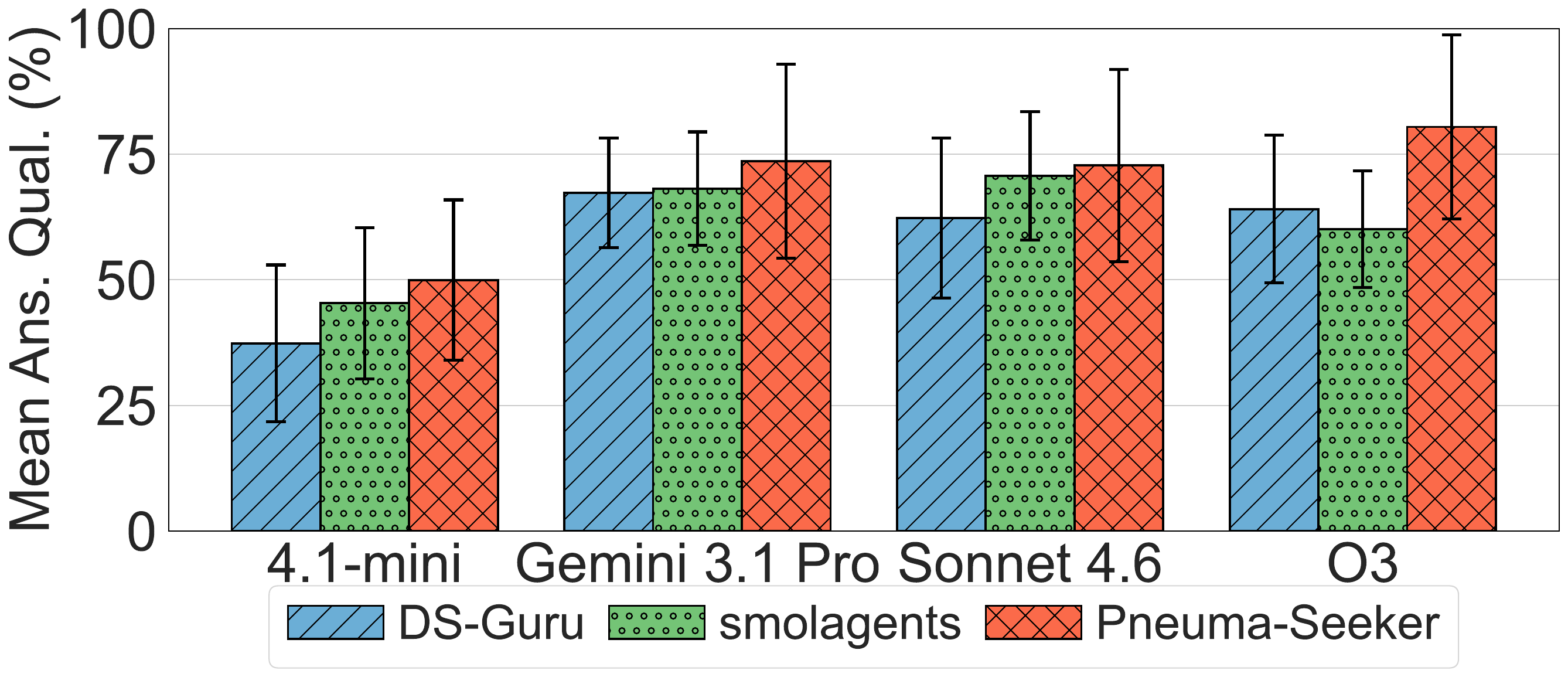}
    \caption{Answer Quality Across Different LLMs}
    \label{fig:llm_swap}
\end{figure}

We compare the systems' answer quality across different LLMs: \texttt{gpt-4.1-mini-2025-} \texttt{04-14}, \texttt{o3-2025-04-16}, \texttt{gemini-3.1-pro-preview}, and \texttt{claude-} \texttt{4.6-sonnet}. We report the mean answer quality and standard deviation across all datasets (besides Astronomy, since \mbox{\dsguru} cannot run on it) in \mbox{Figure~\ref{fig:llm_swap}}. With every LLM, \ps achieves higher answer quality than both baselines. For instance, with \texttt{gemini-3.1-pro-preview}, \mbox{\ps} achieves an answer quality of 73.59\%, which is 5.44 and 5.3 percentage points higher than \mbox{\smolagent} and \mbox{\dsguru}, respectively.

\subsubsection{Memory Usage and Runtime--Memory Trade-offs}
\label{subsubsec:memory_usage_runtime_memory_tradeoffs}
We study the feasibility of running the systems with limited memory (16~GB). Across datasets, \ps's memory usage stays within a narrow range, whereas \dsguru's grows with dataset size.

The largest discrepancy appears on Wildfire: \dsguru loads all rows of all tables into memory and peaks at 5.3~GB, whereas \smolagent and \ps use only 186~MB and 149~MB, respectively. On smaller datasets, the gap narrows or reverses: on Biomedical, \dsguru, \smolagent, and \ps peak at 567, 135, and 186~MB; on Legal, at 25, 43, and 116~MB.

\mypar{Scalability} We also compare memory usage on the scalability dataset (Section~\ref{subsubsec:scalability}). At larger scales, \ps consistently consumes the least memory. For example, at the 1~GB scale, \ps uses 135~MB, whereas \dsguru and \smolagent use 4.4~GB and 916~MB, respectively (Figure~\ref{fig:rq2_scalability_memory}).

\begin{figure}[h]
    \centering
    \includegraphics[width=\linewidth]{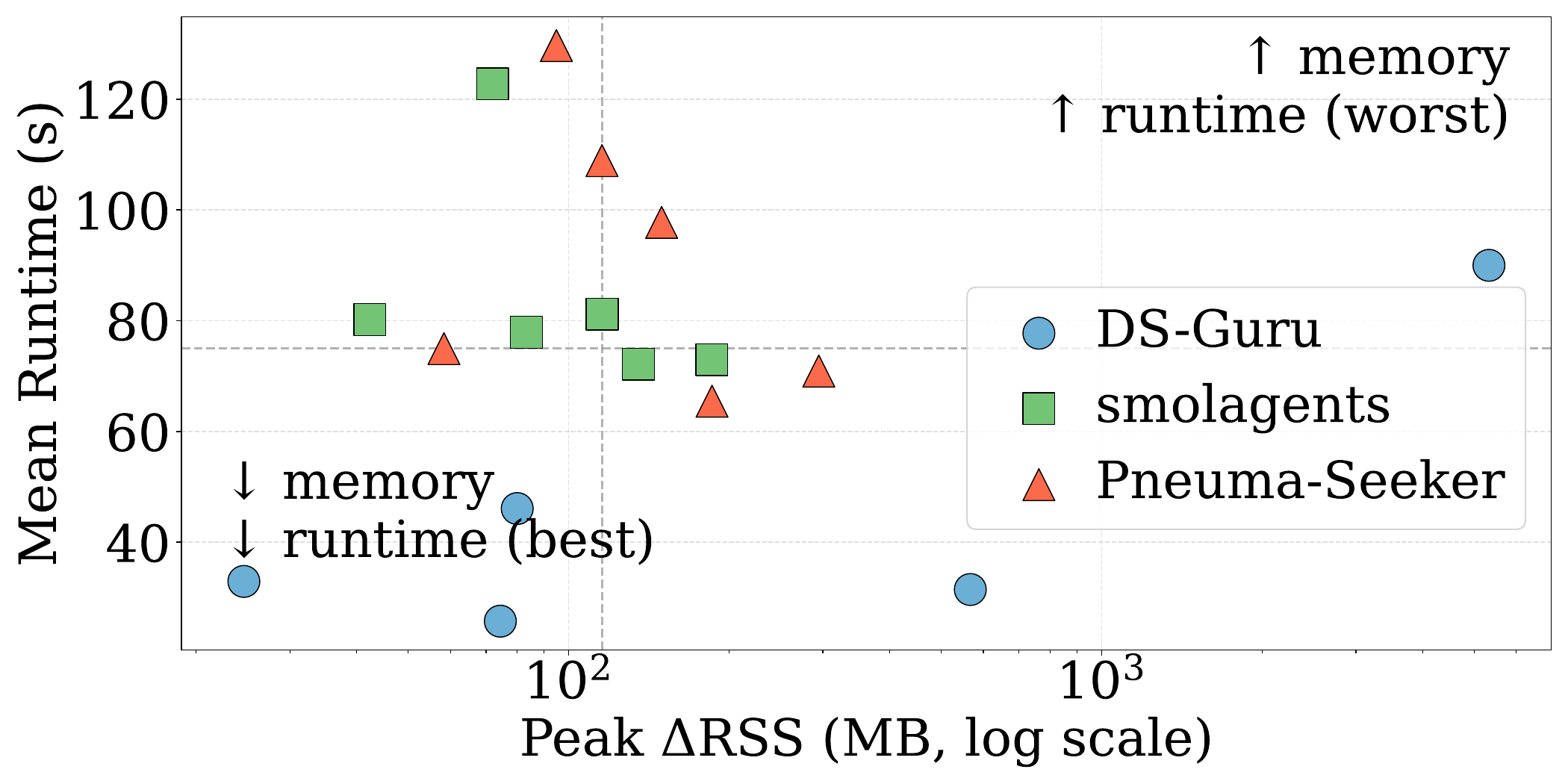}
    \caption{Memory--Runtime Trade-Offs}
    \label{fig:rq2_runtime_memory}
\end{figure}

We observe irregularities in the runtime and memory usage of \smolagent when scaling from 1~GB to 1.9~GB: the reduction at 1.9~GB is due to an internal optimization triggered by data-integrity errors. The smaller dataset was processed using a standard, resource-intensive loading strategy, whereas the larger dataset contained noisy strings that caused the agent to switch to a more selective pipeline, performing column pruning and hash-based filtering with lower memory usage and computational overhead than the generic strategy applied to the smaller dataset.

\mypar{Runtime vs. Memory Usage}
As shown in Figure~\ref{fig:rq2_runtime_memory}, \ps operates in a low-memory, higher-runtime regime. \smolagent is similarly low-memory but on average faster, while \dsguru only minimizes both metrics on small datasets. In short, \ts improves accuracy and explainability (aligning with \textbf{R\ref{req:correctness}} and \textbf{R\ref{req:interpretability}}) for a modest increase in runtime over baselines.

\subsubsection{Retrieval Recall as an Upper Bound on Answer Quality}
\label{subsubsec:retrieval_recall}
The baselines have access to all tables in each dataset, whereas \ps relies on tables retrieved by \pr; its answer quality is therefore upper-bounded by the recall of \pr. We evaluate this recall and compare it against \textsc{Pneuma}~\cite{BalakaPneuma2025} and \textsc{Pneuma} + content search (PCS; Section~\ref{subsec:retriever}), i.e., \pr without table enumeration.

\begin{figure}[h]
    \centering
    \includegraphics[width=\linewidth]{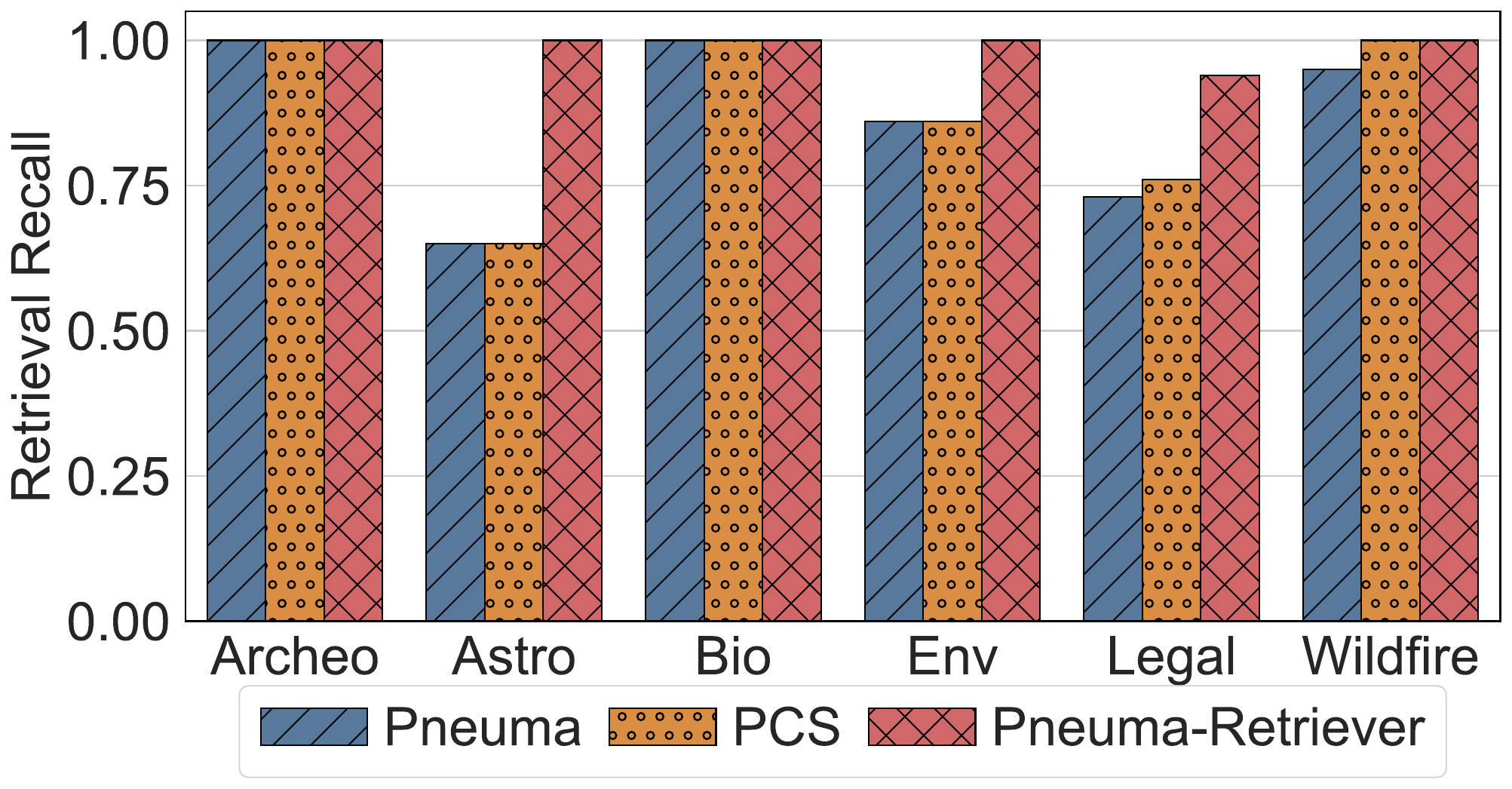}
    \caption{Retrieval Recall}
    \label{fig:rq1_retrieval_recall}
\end{figure}

As shown in Figure~\ref{fig:rq1_retrieval_recall}, PCS slightly improves recall over \textsc{Pneuma} on the Legal and Wildfire datasets, by 3 and 5 percentage points, respectively. In contrast, \pr achieves $100\%$ recall on all datasets except Legal ($94\%$). The recall gap between \pr and the other retrievers is particularly pronounced on the Astronomy, Environment, and Legal datasets.

The primary source of this discrepancy is that \pr allows further enumeration of tables with similar names using regex-based matching, which is not captured by standard top-$k$ retrieval. For example, the query \textit{``Does there exist a metropolitan area in which the number of reports of identity theft exceeded the number of reports of fraud in 2024?''} requires retrieving 104 tables corresponding to fraud and identity-theft reports across all U.S. states. Top-$k$ retrievers systematically under-retrieve in such settings.

%% file: camera-ready-sections/7_related_work.tex
\section{Related Work}
\label{sec:related_work}
We previously introduced \ps in~\cite{BalakaPneumaProject2026}. The system has since evolved along three axes: (i) using Python scripts ($S$) rather than SQL queries ($Q$) for more expressive computation over $\mathcal{T}$; (ii) introducing \dbservice for scalable database-backed actions; and (iii) adding Context Extraction for micro-context management. We next situate \ps with respect to existing data systems and related research.

\mypar{Enterprise Data Systems}
Existing data systems partially address the problem: business intelligence tools (e.g., \mbox{\textsc{Tableau}~\cite{tableau}} or \textsc{Power}~\mbox{BI~\cite{powerbi}}) support interactive analysis, data \mbox{catalogs~\cite{JahnkeDataCatalogs2023}} (e.g., \mbox{\textsc{Goods}~\cite{Goods}} and \mbox{\textsc{Amundsen}~\cite{amundsen}}) improve dataset discovery and documentation, and data integration platforms (e.g., \mbox{\textsc{Fivetran}~\cite{fivetran}}) automate data ingestion. Yet these systems largely require users to translate high-level analytical tasks into concrete data requirements, discover relevant tables, and iteratively refine \mbox{\activeInfoNeed} through manual exploration, so moving from \mbox{\activeInfoNeed} to \mbox{\latentInfoNeed} remains time-consuming and expertise-dependent, motivating interactive systems for iterative refinement of evolving information needs.

\mypar{Sensemaking and Exploratory Search}
Both frame analysis as iterative processes of foraging, hypothesis formation, and synthesis with feedback and intermediate representations~\cite{pirolli2005sensemaking,GaryExploratorySearch2006,HearstSearchUI2009}, motivating \ps's framing of tabular QA as interactive sensemaking rather than one-shot query answering. Recent mixed-initiative systems such as \textsc{DocWrangler}~\cite{DocWrangler2025}, \textsc{ScholarMate}~\cite{YeScholarMate2025}, and \textsc{Sensecape}~\cite{SuhSensecape2023} similarly support iterative refinement, inspection, and evolving information needs.


\mypar{Agentic Systems for Tabular Question Answering (QA)}
Recent LLM-based systems move beyond single-step retrieval or end-to-end generation (e.g., RAG~\cite{LewisRAG2020}) toward agentic architectures~\cite{yue2025surveylargelanguagemodel} that implement reason--act--reflect loops (e.g., ReAct~\cite{yao2023reactsynergizingreasoningacting}) and orchestrate tools such as database queries, code execution, and web search~\cite{CoordinatedLLMAgents2025,wang2025lmarslegalmultiagentworkflow}. For Tabular QA, \textsc{ReAcTable}~\cite{ReactTable2024}, \textsc{Chain-of-Table}~\cite{wang2024chainoftable}, and \textsc{TableZoomer}~\cite{XiongTableZoomer2025} leverage intermediate programs and iterative validation, but assume small curated table sets or single-table reasoning. In contrast, \ps couples agentic reasoning with data discovery to support multi-table reasoning.

Industrial systems such as \mbox{\textsc{Databricks Genie}~\cite{databricks_genie_2026}} map natural-language questions to SQL queries and resolve ambiguity through LLM-generated clarifying questions. \mbox{\ps} similarly asks clarifying questions, but also exposes a sidebar for inspection.

\mypar{Tabular Data Discovery}
Tabular data discovery spans metadata-driven search, view identification, schema matching, and natural-language interfaces. Some systems emphasize structured discovery and joinability analysis, such as \textsc{Aurum}~\cite{Aurum2018}, \textsc{Auctus}~\cite{Auctus2021}, \textsc{Ver}~\cite{GongVer2023}, and FREYJA~\cite{maynou2026freyja}. Others leverage learned representations for natural-language and multi-table discovery, such as \textsc{Solo}~\cite{Solo2024} and \textsc{Octopus}~\cite{LiOctopus2026}. Our prior system, \pneuma~\cite{BalakaPneuma2025}, underlies the discovery component in \ps; here we extend it to improve recall beyond top-$k$ to support multi-table reasoning, as downstream agentic performance is bounded by discovery completeness.

%% file: camera-ready-sections/8_conclusion.tex
\section{Conclusion}
\label{sec:conclusion}

This paper presented relational reification and a system, \ps, that implements this idea within an LLM-powered agentic system to help users articulate their information needs and answer them. Our results show that representing evolving intent as an explicit \ts pair improves answer quality while enabling inspectability. Taken together, these findings suggest that relational reification is a practical foundation for trustworthy LLM-mediated data discovery and preparation.

%% file: camera-ready-sections/9_acknowledgement.tex
\begin{acks}
    We would like to thank the Alfred P. Sloan Foundation, NSF (Award \#2340034), and the Data Ecology Research Initiative at the Data Science Institute at The University of Chicago for supporting this work.
\end{acks}

\balance

%% file: main.bib
@article{BalakaPneuma2025,
    author = {Balaka, Muhammad Imam Luthfi and Alexander, David and Wang, Qiming and Gong, Yue and Krisnadhi, Adila and Castro Fernandez, Raul},
    title = {Pneuma: Leveraging LLMs for Tabular Data Representation and Retrieval in an End-to-End System},
    year = {2025},
    issue_date = {June 2025},
    publisher = {Association for Computing Machinery},
    address = {New York, NY, USA},
    volume = {3},
    number = {3},
    url = {https://doi.org/10.1145/3725337},
    doi = {10.1145/3725337},
    journal = {Proc. ACM Manag. Data},
    month = jun,
    articleno = {200},
    numpages = {28}
}

@inproceedings{RaasveldtDuckDB2019,
author = {Raasveldt, Mark and M\"{u}hleisen, Hannes},
title = {DuckDB: an Embeddable Analytical Database},
year = {2019},
isbn = {9781450356435},
publisher = {Association for Computing Machinery},
address = {New York, NY, USA},
url = {https://doi.org/10.1145/3299869.3320212},
doi = {10.1145/3299869.3320212},
booktitle = {Proceedings of the 2019 International Conference on Management of Data},
pages = {1981–1984},
numpages = {4},
location = {Amsterdam, Netherlands},
series = {SIGMOD '19}
}

@misc{LiOctopus2026,
title={Octopus: A Lightweight Entity-Aware System for Multi-Table Data Discovery and Cell-Level Retrieval}, 
author={Wen-Zhi Li and Sainyam Galhotra},
year={2026},
eprint={2601.02304},
archivePrefix={arXiv},
primaryClass={cs.DB},
url={https://arxiv.org/abs/2601.02304}, 
}

@inproceedings{LiuPalimpzest2025,
    title={Palimpzest: Optimizing AI-Powered Analytics with Declarative Query Processing},
    author={Liu, Chunwei and Russo, Matthew and Cafarella, Michael and Cao, Lei and Chen, Peter Baile and Chen, Zui and Franklin, Michael and Kraska, Tim and Madden, Samuel and Shahout, Rana and Vitagliano, Gerardo},
    booktitle = {Proceedings of the {{Conference}} on {{Innovative Database Research}} ({{CIDR}})},
    date = 2025,
}

@inproceedings{BalakaPneumaProject2026,
  author       = {Muhammad Imam Luthfi Balaka and Raul Castro Fernandez},
  title        = {The Pneuma Project: Reifying Information Needs as Relational Schemas to Automate Discovery, Guide Preparation, and Align Data with Intent},
  booktitle = {Proceedings of the {{Conference}} on {{Innovative Database Research}} ({{CIDR}})},
  year         = {2026},
}

@INPROCEEDINGS{Aurum2018,
  author={Castro Fernandez, Raul and Abedjan, Ziawasch and Koko, Famien and Yuan, Gina and Madden, Samuel and Stonebraker, Michael},
  booktitle={2018 IEEE 34th International Conference on Data Engineering (ICDE)}, 
  title={Aurum: A Data Discovery System}, 
  year={2018},
  volume={},
  number={},
  pages={1001-1012},
  doi={10.1109/ICDE.2018.00094}
}

@misc{wang2025lmarslegalmultiagentworkflow,
      title={L-MARS: Legal Multi-Agent Workflow with Orchestrated Reasoning and Agentic Search}, 
      author={Ziqi Wang and Boqin Yuan},
      year={2025},
      eprint={2509.00761},
      archivePrefix={arXiv},
      primaryClass={cs.AI},
      url={https://arxiv.org/abs/2509.00761}, 
}

@INPROCEEDINGS{GongVer2023,
  author={Gong, Yue and Zhu, Zhiru and Galhotra, Sainyam and Fernandez, Raul Castro},
  booktitle={2023 IEEE 39th International Conference on Data Engineering (ICDE)}, 
  title={Ver: View Discovery in the Wild}, 
  year={2023},
  volume={},
  number={},
  pages={503-516},
  doi={10.1109/ICDE55515.2023.00045}}

@article{Auctus2021,
author = {Castelo, Sonia and Rampin, R\'{e}mi and Santos, A\'{e}cio and Bessa, Aline and Chirigati, Fernando and Freire, Juliana},
title = {Auctus: a dataset search engine for data discovery and augmentation},
year = {2021},
issue_date = {July 2021},
publisher = {VLDB Endowment},
volume = {14},
number = {12},
issn = {2150-8097},
url = {https://doi.org/10.14778/3476311.3476346},
doi = {10.14778/3476311.3476346},
journal = {Proc. VLDB Endow.},
month = jul,
pages = {2791–2794},
numpages = {4}
}

@inproceedings{jin2026Text2SQLBroken,
  title        = {Text-to-SQL Benchmarks are Broken: An In-Depth Analysis of Annotation Errors},
  author       = {Tengjun Jin and Yoojin Choi and Yuxuan Zhu and Daniel Kang},
  booktitle    = {Proceedings of the 16th Annual Conference on Innovative Data Systems Research (CIDR '26)},
  year         = {2026},
  address      = {Chaminade, USA},
  month        = {January 18--21},
  url          = {https://vldb.org/cidrdb/2026/text-to-sql-benchmarks-are-broken-an-in-depth-analysis-of-annotation-errors.html}
}

@Article{XiongTableZoomer2025,
author={Xiong, Sishi
and He, Ziyang
and He, Zhongjiang
and Zhao, Yu
and Pan, Changzai
and Zhang, Jie
and Song, Shuangyong
and Li, Yongxiang},
title={TableZoomer: a collaborative agent framework for large-scale table question answering},
journal={Vicinagearth},
year={2025},
month={Nov},
day={06},
volume={2},
number={1},
pages={11},
issn={3005-060X},
doi={10.1007/s44336-025-00016-x},
url={https://doi.org/10.1007/s44336-025-00016-x}
}

@article{CoordinatedLLMAgents2025,
title = {Coordinated LLM multi-agent systems for collaborative question-answer generation},
journal = {Knowledge-Based Systems},
volume = {330},
pages = {114627},
year = {2025},
issn = {0950-7051},
doi = {https://doi.org/10.1016/j.knosys.2025.114627},
url = {https://www.sciencedirect.com/science/article/pii/S0950705125016661},
author = {Sami Saadaoui and Eduardo Alonso},
}

@misc{yao2023reactsynergizingreasoningacting,
      title={ReAct: Synergizing Reasoning and Acting in Language Models}, 
      author={Shunyu Yao and Jeffrey Zhao and Dian Yu and Nan Du and Izhak Shafran and Karthik Narasimhan and Yuan Cao},
      year={2023},
      eprint={2210.03629},
      archivePrefix={arXiv},
      primaryClass={cs.CL},
      url={https://arxiv.org/abs/2210.03629}, 
}

@Article{NumajiriOpenData2024,
author={Numajiri, Honami
and Hayashi, Takayuki},
title={Analysis on open data as a foundation for data-driven research},
journal={Scientometrics},
year={2024},
month={Oct},
day={01},
volume={129},
number={10},
pages={6315-6332},
issn={1588-2861},
doi={10.1007/s11192-024-04956-x},
url={https://doi.org/10.1007/s11192-024-04956-x}
}

@article{PredictingCustomerPurchase2024,
title = {Predicting online customer purchase: The integration of customer characteristics and browsing patterns},
journal = {Decision Support Systems},
volume = {177},
pages = {114105},
year = {2024},
issn = {0167-9236},
doi = {https://doi.org/10.1016/j.dss.2023.114105},
url = {https://www.sciencedirect.com/science/article/pii/S016792362300180X},
author = {Seongbeom Kim and Woosik Shin and Hee-Woong Kim},
}

@inproceedings{pirolli2005sensemaking,
  title        = {The Sensemaking Process and Leverage Points for Analyst Technology as Identified Through Cognitive Task Analysis},
  author       = {Pirolli, Peter and Card, Stuart},
  booktitle    = {Proceedings of the International Conference on Intelligence Analysis},
  year         = {2005},
  address      = {McLean, VA},
  publisher    = {MITRE},
}

@book{HearstSearchUI2009,
place={Cambridge},
title={Search User Interfaces},
publisher={Cambridge University Press},
author={Hearst, Marti A.},
year={2009}}

@misc{yue2025surveylargelanguagemodel,
      title={A Survey of Large Language Model Agents for Question Answering}, 
      author={Murong Yue},
      year={2025},
      eprint={2503.19213},
      archivePrefix={arXiv},
      primaryClass={cs.CL},
      url={https://arxiv.org/abs/2503.19213}, 
}

@inproceedings{LewisRAG2020,
author = {Lewis, Patrick and Perez, Ethan and Piktus, Aleksandra and Petroni, Fabio and Karpukhin, Vladimir and Goyal, Naman and K\"{u}ttler, Heinrich and Lewis, Mike and Yih, Wen-tau and Rockt\"{a}schel, Tim and Riedel, Sebastian and Kiela, Douwe},
title = {Retrieval-augmented generation for knowledge-intensive NLP tasks},
year = {2020},
isbn = {9781713829546},
publisher = {Curran Associates Inc.},
address = {Red Hook, NY, USA},
booktitle = {Proceedings of the 34th International Conference on Neural Information Processing Systems},
articleno = {793},
numpages = {16},
location = {Vancouver, BC, Canada},
series = {NIPS '20}
}

@inproceedings{SuhSensecape2023,
author = {Suh, Sangho and Min, Bryan and Palani, Srishti and Xia, Haijun},
title = {Sensecape: Enabling Multilevel Exploration and Sensemaking with Large Language Models},
year = {2023},
isbn = {9798400701320},
publisher = {Association for Computing Machinery},
address = {New York, NY, USA},
url = {https://doi.org/10.1145/3586183.3606756},
doi = {10.1145/3586183.3606756},
booktitle = {Proceedings of the 36th Annual ACM Symposium on User Interface Software and Technology},
articleno = {1},
numpages = {18},
location = {San Francisco, CA, USA},
series = {UIST '23}
}

@inproceedings{YeScholarMate2025,
author = {Ye, Runlong and Lee, Patrick and Varona, Matthew and Huang, Oliver and Nobre, Carolina},
title = {ScholarMate: A Mixed-Initiative Tool for Qualitative Knowledge Work and Information Sensemaking},
year = {2025},
isbn = {9798400713972},
publisher = {Association for Computing Machinery},
address = {New York, NY, USA},
url = {https://doi.org/10.1145/3707640.3731913},
doi = {10.1145/3707640.3731913},
booktitle = {Adjunct Proceedings of the 4th Annual Symposium on Human-Computer Interaction for Work},
articleno = {7},
numpages = {7},
location = {
},
series = {CHIWORK '25 Adjunct}
}

@article{GaryExploratorySearch2006,
author = {Marchionini, Gary},
title = {Exploratory search: from finding to understanding},
year = {2006},
issue_date = {April 2006},
publisher = {Association for Computing Machinery},
address = {New York, NY, USA},
volume = {49},
number = {4},
issn = {0001-0782},
url = {https://doi.org/10.1145/1121949.1121979},
doi = {10.1145/1121949.1121979},
journal = {Commun. ACM},
month = apr,
pages = {41–46},
numpages = {6}
}

@misc{smolagents2024,
  author = {Hugging Face},
  title = {smolagents: a lightweight library to build agents that write and run Python code},
  year = {2024},
  publisher = {GitHub},
  howpublished = {\url{https://github.com/huggingface/smolagents}},
  note = {Accessed: 2026-02-28}
}

@ARTICLE{DataLakes2023,
  author={Hai, Rihan and Koutras, Christos and Quix, Christoph and Jarke, Matthias},
  journal={IEEE Transactions on Knowledge and Data Engineering}, 
  title={Data Lakes: A Survey of Functions and Systems}, 
  year={2023},
  volume={35},
  number={12},
  pages={12571-12590},
  doi={10.1109/TKDE.2023.3270101}}

@article{DataDiscoveryInDataLakes2025,
author = {Abedjan, Ziawasch and Esmailoghli, Mahdi and Galhotra, Sainyam},
title = {Data Discovery in Data Lakes: Operations, Indexes, Systems},
year = {2025},
issue_date = {August 2025},
publisher = {VLDB Endowment},
volume = {18},
number = {12},
issn = {2150-8097},
url = {https://doi.org/10.14778/3750601.3750694},
doi = {10.14778/3750601.3750694},
journal = {Proc. VLDB Endow.},
month = aug,
pages = {5455–5459},
numpages = {5}
}

@inproceedings{LLMZeroShotReasoners2022,
author = {Kojima, Takeshi and Gu, Shixiang Shane and Reid, Machel and Matsuo, Yutaka and Iwasawa, Yusuke},
title = {Large language models are zero-shot reasoners},
year = {2022},
isbn = {9781713871088},
publisher = {Curran Associates Inc.},
address = {Red Hook, NY, USA},
booktitle = {Proceedings of the 36th International Conference on Neural Information Processing Systems},
articleno = {1613},
numpages = {15},
location = {New Orleans, LA, USA},
series = {NIPS '22}
}

@inproceedings{CoT2022,
author = {Wei, Jason and Wang, Xuezhi and Schuurmans, Dale and Bosma, Maarten and Ichter, Brian and Xia, Fei and Chi, Ed H. and Le, Quoc V. and Zhou, Denny},
title = {Chain-of-thought prompting elicits reasoning in large language models},
year = {2022},
isbn = {9781713871088},
publisher = {Curran Associates Inc.},
address = {Red Hook, NY, USA},
booktitle = {Proceedings of the 36th International Conference on Neural Information Processing Systems},
articleno = {1800},
numpages = {14},
location = {New Orleans, LA, USA},
series = {NIPS '22}
}

@Article{JahnkeDataCatalogs2023,
author={Jahnke, Nils
and Otto, Boris},
title={Data Catalogs in the Enterprise: Applications and Integration},
journal={Datenbank-Spektrum},
year={2023},
month={Jul},
day={01},
volume={23},
number={2},
pages={89-96},
issn={1610-1995},
doi={10.1007/s13222-023-00445-2},
url={https://doi.org/10.1007/s13222-023-00445-2}
}

@article{Goods,title	= {Goods: Organizing Google's Datasets},author	= {Alon Halevy and Flip Korn and Natalya F. Noy and Christopher Olston and Neoklis Polyzotis and Sudip Roy and Steven Euijong Whang},year	= {2016},journal	= {SIGMOD}}

@article{likert1932technique,
  title={A technique for the measurement of attitudes},
  author={Likert, Rensis},
  journal={Archives of Psychology},
  volume={22},
  number={140},
  pages={1--55},
  year={1932},
  publisher={Archives of Psychology}
}

@article{2024LiuLostInTheMiddle,
    title = "Lost in the Middle: How Language Models Use Long Contexts",
    author = "Liu, Nelson F.  and
      Lin, Kevin  and
      Hewitt, John  and
      Paranjape, Ashwin  and
      Bevilacqua, Michele  and
      Petroni, Fabio  and
      Liang, Percy",
    journal = "Transactions of the Association for Computational Linguistics",
    volume = "12",
    year = "2024",
    address = "Cambridge, MA",
    publisher = "MIT Press",
    url = "https://aclanthology.org/2024.tacl-1.9/",
    doi = "10.1162/tacl_a_00638",
    pages = "157--173",
}

@misc{databricks_genie_2026,
  title        = {The next generation of Databricks Genie},
  author       = {{Databricks}},
  year         = {2026},
  month        = apr,
  url          = {https://www.databricks.com/blog/next-generation-databricks-genie},
  note         = {Accessed: 2026-05-24}
}

@misc{tableau,
  title        = {Tableau Business Intelligence and Analytics Software},
  author       = {{Tableau}},
  year         = {2026},
  month        = june,
  url          = {https://www.tableau.com},
  note         = {Accessed: 2026-06-01}
}

@misc{powerbi,
  title        = {Power BI: Data Visualization Software},
  author       = {{Microsoft}},
  year         = {2026},
  month        = june,
  url          = {https://powerbi.microsoft.com},
  note         = {Accessed: 2026-06-01}
}

@misc{amundsen,
  title        = {Amundsen: Open-source Data Discovery and Metadata Engine},
  author       = {{Amundsen}},
  year         = {2026},
  month        = june,
  url          = {https://www.amundsen.io},
  note         = {Accessed: 2026-06-01}
}

@misc{fivetran,
  title        = {Fivetran: Automated Data Movement Platform},
  author       = {{Fivetran}},
  year         = {2026},
  month        = june,
  url          = {https://www.fivetran.com},
  note         = {Accessed: 2026-06-01}
}

@article{vod,
    author = {Castro Fernandez, Raul},
    title = {What is the Value of Data? A Theory and Systematization},
    year = {2025},
    issue_date = {March 2025},
    publisher = {Association for Computing Machinery},
    address = {New York, NY, USA},
    volume = {2},
    number = {1},
    url = {https://doi.org/10.1145/3728476},
    doi = {10.1145/3728476},
    journal = {ACM / IMS J. Data Sci.},
    month = jun,
    articleno = {3},
    numpages = {25}
}

@article{1996ScaifeExternalCognition,
title = {External cognition: how do graphical representations work?},
journal = {International Journal of Human-Computer Studies},
volume = {45},
number = {2},
pages = {185-213},
year = {1996},
issn = {1071-5819},
doi = {https://doi.org/10.1006/ijhc.1996.0048},
url = {https://www.sciencedirect.com/science/article/pii/S1071581996900488},
author = {Mike Scaife and Yvonne Rogers},
}

@article{Belkin:1980,
  author = {Belkin, N.J.},
  title = {Anomalous states of knowledge as a basis for information retrieval},
  journal = {Canadian Journal of Information Science},
  volume = {5},
  number = {1},
  pages = {133--143},
  year = {1980}
}

@inproceedings{PatelSemanticOperator2024,
  title={Semantic Operators: A Declarative Model for Rich, AI-based Data Processing},
  author={Liana Patel and Siddharth Jha and Melissa Pan and Harshit Gupta and Parth Asawa and Carlos Guestrin and Matei Zaharia},
  year={2024},
  url={https://api.semanticscholar.org/CorpusID:271218837}
}

@article{maynou2026freyja,
  title={FREYJA: Efficient join discovery in data lakes},
  author={Maynou, Marc and Nadal, Sergi and Panadero, Raquel and Flores, Javier and Romero, Oscar and Queralt, Anna},
  journal={IEEE Transactions on Knowledge and Data Engineering},
  year={2026},
  publisher={IEEE}
}

@INPROCEEDINGS{SeepingSemantics2018,
  author={Castro Fernandez, Raul and Mansour, Essam and Qahtan, Abdulhakim A. and Elmagarmid, Ahmed and Ilyas, Ihab and Madden, Samuel and Ouzzani, Mourad and Stonebraker, Michael and Tang, Nan},
  booktitle={2018 IEEE 34th International Conference on Data Engineering (ICDE)}, 
  title={Seeping Semantics: Linking Datasets Using Word Embeddings for Data Discovery}, 
  year={2018},
  volume={},
  number={},
  pages={989-1000},
  doi={10.1109/ICDE.2018.00093}}

@misc{xing2025MMTU,
  title        = {MMTU: A Massive Multi-Task Table Understanding and Reasoning Benchmark},
  author       = {Junjie Xing and Yeye He and Mengyu Zhou and Haoyu Dong and Shi Han and Lingjiao Chen and Dongmei Zhang and Surajit Chaudhuri and H. V. Jagadish},
  year         = {2025},
  eprint       = {2506.05587},
  archivePrefix= {arXiv},
  primaryClass = {cs.AI},
  url          = {https://arxiv.org/abs/2506.05587},
}

@inproceedings{liu-etal-2025-user,
    title = "User Feedback in Human-{LLM} Dialogues: A Lens to Understand Users But Noisy as a Learning Signal",
    author = "Liu, Yuhan  and
      Zhang, Michael JQ  and
      Choi, Eunsol",
    editor = "Christodoulopoulos, Christos  and
      Chakraborty, Tanmoy  and
      Rose, Carolyn  and
      Peng, Violet",
    booktitle = "Proceedings of the 2025 Conference on Empirical Methods in Natural Language Processing",
    month = nov,
    year = "2025",
    address = "Suzhou, China",
    publisher = "Association for Computational Linguistics",
    url = "https://aclanthology.org/2025.emnlp-main.133/",
    doi = "10.18653/v1/2025.emnlp-main.133",
    pages = "2666--2681",
    ISBN = "979-8-89176-332-6",
}

@inproceedings{CaptureKnowledgeUrban2025,
author = {Urban, Matthias and Ding, Jialin and Kernert, David and Vaidya, Kapil and Kraska, Tim},
title = {Utilizing Past User Feedback for More Accurate Text-to-SQL},
year = {2025},
isbn = {9798400719592},
publisher = {Association for Computing Machinery},
address = {New York, NY, USA},
url = {https://doi.org/10.1145/3736733.3736739},
doi = {10.1145/3736733.3736739},
booktitle = {Proceedings of the Workshop on Human-In-the-Loop Data Analytics},
articleno = {10},
numpages = {7},
location = {Intercontinental Berlin, Berlin, Germany},
series = {HILDA '25}
}

@article{ReactTable2024,
author = {Zhang, Yunjia and Henkel, Jordan and Floratou, Avrilia and Cahoon, Joyce and Deep, Shaleen and Patel, Jignesh M.},
title = {ReAcTable: Enhancing ReAct for Table Question Answering},
year = {2024},
issue_date = {April 2024},
publisher = {VLDB Endowment},
volume = {17},
number = {8},
issn = {2150-8097},
url = {https://doi.org/10.14778/3659437.3659452},
doi = {10.14778/3659437.3659452},
journal = {Proc. VLDB Endow.},
month = apr,
pages = {1981–1994},
numpages = {14}
}

@inproceedings{
wang2024chainoftable,
title={Chain-of-Table: Evolving Tables in the Reasoning Chain for Table Understanding},
author={Zilong Wang and Hao Zhang and Chun-Liang Li and Julian Martin Eisenschlos and Vincent Perot and Zifeng Wang and Lesly Miculicich and Yasuhisa Fujii and Jingbo Shang and Chen-Yu Lee and Tomas Pfister},
booktitle={The Twelfth International Conference on Learning Representations},
year={2024},
url={https://openreview.net/forum?id=4L0xnS4GQM}
}

@inproceedings{DocWrangler2025,
author = {Shankar, Shreya and Chopra, Bhavya and Hasan, Mawil and Lee, Stephen and Hartmann, Bjoern and Hellerstein, Joseph and Parameswaran, Aditya and Wu, Eugene},
title = {Steering Semantic Data Processing With DocWrangler},
year = {2025},
isbn = {9798400720376},
publisher = {Association for Computing Machinery},
address = {New York, NY, USA},
url = {https://doi.org/10.1145/3746059.3747625},
doi = {10.1145/3746059.3747625},
booktitle = {Proceedings of the 38th Annual ACM Symposium on User Interface Software and Technology},
articleno = {84},
numpages = {18},
location = {
},
series = {UIST '25}
}

@inproceedings{park-etal-2021-scalable,
    title = "A Scalable Framework for Learning From Implicit User Feedback to Improve Natural Language Understanding in Large-Scale Conversational {AI} Systems",
    author = "Park, Sunghyun  and
      Li, Han  and
      Patel, Ameen  and
      Mudgal, Sidharth  and
      Lee, Sungjin  and
      Kim, Young-Bum  and
      Matsoukas, Spyros  and
      Sarikaya, Ruhi",
    editor = "Moens, Marie-Francine  and
      Huang, Xuanjing  and
      Specia, Lucia  and
      Yih, Scott Wen-tau",
    booktitle = "Proceedings of the 2021 Conference on Empirical Methods in Natural Language Processing",
    month = nov,
    year = "2021",
    address = "Online and Punta Cana, Dominican Republic",
    publisher = "Association for Computational Linguistics",
    url = "https://aclanthology.org/2021.emnlp-main.489/",
    doi = "10.18653/v1/2021.emnlp-main.489",
    pages = "6054--6063",
}

@inproceedings{li2024BIRD,
  title={Can LLM already serve as a database interface? A big bench for large-scale database grounded text-to-SQLs},
  author={Li, Shiyao and Ning, Xuefei and Wang, Luning and Liu, Tengxuan and Shi, Xiangsheng and Yan, Shengen and Dai, Guohao and Yang, Huazhong and Wang, Yu},
  booktitle={Proceedings of the 37th International Conference on Neural Information Processing Systems (NIPS '23)},
  year={2024},
  organization={Curran Associates, Inc.},
  doi={10.5555/3666122.3667957}
}

@misc{biswal2024TAG,
      title={Text2SQL is Not Enough: Unifying AI and Databases with TAG}, 
      author={Asim Biswal and Liana Patel and Siddarth Jha and Amog Kamsetty and Shu Liu and Joseph E. Gonzalez and Carlos Guestrin and Matei Zaharia},
      year={2024},
      eprint={2408.14717},
      archivePrefix={arXiv},
      primaryClass={cs.DB},
      url={https://arxiv.org/abs/2408.14717}, 
}

@inproceedings{pal2023multitabqa,
    title = "{M}ulti{T}ab{QA}: Generating Tabular Answers for Multi-Table Question Answering",
    author = "Pal, Vaishali  and
      Yates, Andrew  and
      Kanoulas, Evangelos  and
      de Rijke, Maarten",
    editor = "Rogers, Anna  and
      Boyd-Graber, Jordan  and
      Okazaki, Naoaki",
    booktitle = "Proceedings of the 61st Annual Meeting of the Association for Computational Linguistics (Volume 1: Long Papers)",
    month = jul,
    year = "2023",
    address = "Toronto, Canada",
    publisher = "Association for Computational Linguistics",
    url = "https://aclanthology.org/2023.acl-long.348/",
    doi = "10.18653/v1/2023.acl-long.348",
    pages = "6322--6334",
}

@inproceedings{oses2024databench,
    title = "Question Answering over Tabular Data with {D}ata{B}ench: A Large-Scale Empirical Evaluation of {LLM}s",
    author = "Os{\'e}s Grijalba, Jorge  and
      Ure{\~n}a-L{\'o}pez, L. Alfonso  and
      Mart{\'i}nez C{\'a}mara, Eugenio  and
      Camacho-Collados, Jose",
    editor = "Calzolari, Nicoletta  and
      Kan, Min-Yen  and
      Hoste, Veronique  and
      Lenci, Alessandro  and
      Sakti, Sakriani  and
      Xue, Nianwen",
    booktitle = "Proceedings of the 2024 Joint International Conference on Computational Linguistics, Language Resources and Evaluation (LREC-COLING 2024)",
    month = may,
    year = "2024",
    address = "Torino, Italia",
    publisher = "ELRA and ICCL",
    url = "https://aclanthology.org/2024.lrec-main.1179/",
    pages = "13471--13488",
}

@misc{gu2025RADAR,
      title={RADAR: Benchmarking Language Models on Imperfect Tabular Data}, 
      author={Ken Gu and Zhihan Zhang and Kate Lin and Yuwei Zhang and Akshay Paruchuri and Hong Yu and Mehran Kazemi and Kumar Ayush and A. Ali Heydari and Maxwell A. Xu and Girish Narayanswamy and Yun Liu and Ming-Zher Poh and Yuzhe Yang and Mark Malhotra and Shwetak Patel and Hamid Palangi and Xuhai Xu and Daniel McDuff and Tim Althoff and Xin Liu},
      year={2025},
      eprint={2506.08249},
      archivePrefix={arXiv},
      primaryClass={cs.DB},
      url={https://arxiv.org/abs/2506.08249}, 
}

@misc{lai2025kramabenchbenchmarkaisystems,
      title={KramaBench: A Benchmark for AI Systems on Data-to-Insight Pipelines over Data Lakes}, 
      author={Eugenie Lai and Gerardo Vitagliano and Ziyu Zhang and Om Chabra and Sivaprasad Sudhir and Anna Zeng and Anton A. Zabreyko and Chenning Li and Ferdi Kossmann and Jialin Ding and Jun Chen and Markos Markakis and Matthew Russo and Weiyang Wang and Ziniu Wu and Michael J. Cafarella and Lei Cao and Samuel Madden and Tim Kraska},
      year={2025},
      eprint={2506.06541},
      archivePrefix={arXiv},
      primaryClass={cs.DB},
      url={https://arxiv.org/abs/2506.06541}, 
}

@inproceedings{FernandezDataDI2025,
  author    = {Raul Castro Fernandez},
  title     = {Data Discovery is a Socio‑Technical Problem: the Path from Document Identification and Retrieval to Data Ecology},
  booktitle = {IEEE Computer Society Data Engineering Bulletin},
  year      = {2025},
  note      = {Preprint available at Semantic Scholar (CorpusID:281957015)},
  url       = {https://api.semanticscholar.org/CorpusID:281957015}
}

@inproceedings{context-length-hurts-llm,
    title = "Context Length Alone Hurts {LLM} Performance Despite Perfect Retrieval",
    author = "Du, Yufeng  and
      Tian, Minyang  and
      Ronanki, Srikanth  and
      Rongali, Subendhu  and
      Bodapati, Sravan Babu  and
      Galstyan, Aram  and
      Wells, Azton  and
      Schwartz, Roy  and
      Huerta, Eliu A  and
      Peng, Hao",
    editor = "Christodoulopoulos, Christos  and
      Chakraborty, Tanmoy  and
      Rose, Carolyn  and
      Peng, Violet",
    booktitle = "Findings of the Association for Computational Linguistics: EMNLP 2025",
    month = nov,
    year = "2025",
    address = "Suzhou, China",
    publisher = "Association for Computational Linguistics",
    url = "https://aclanthology.org/2025.findings-emnlp.1264/",
    doi = "10.18653/v1/2025.findings-emnlp.1264",
    pages = "23281--23298",
    ISBN = "979-8-89176-335-7",
}

@article{Solo2024,
author = {Wang, Qiming and Castro Fernandez, Raul},
title = {Solo: Data Discovery Using Natural Language Questions Via A Self-Supervised Approach},
year = {2023},
issue_date = {December 2023},
publisher = {Association for Computing Machinery},
address = {New York, NY, USA},
volume = {1},
number = {4},
url = {https://doi.org/10.1145/3626756},
doi = {10.1145/3626756},
journal = {Proc. ACM Manag. Data},
month = dec,
articleno = {262},
numpages = {27}
}
